%% file: blackhole-paper.tex
\documentclass[12pt]{article}
\usepackage[margin=1in]{geometry}
\usepackage[pdftex,colorlinks=true]{hyperref}  
\usepackage{array}
\usepackage{booktabs}
\usepackage{amsmath}
\usepackage{amsthm}
\usepackage{amssymb}
\usepackage{bm}
\usepackage{mathrsfs}
\usepackage{graphicx}
\usepackage{color}

\numberwithin{equation}{section}

\input{Ricci}
\DeclareMathOperator{\so}{\mathfrak{s}\mathfrak{o}}

\DeclareMathOperator{\SOrth}{SO}

\DeclareMathOperator{\Orthframe}{\mathcal{F}}
\DeclareMathOperator{\Redframe}{\Orthframe_{\text{red}}}
\DeclareMathOperator{\Lieder}{\mathscr{L}}

\newcommand{\RS}{R_{S}}
\newcommand{\bnabla}{\bm{\nabla}}
\newcommand{\orbit}{\mathcal{O}}

\newcommand{\bbR}{\mathbb{R}}
\newcommand{\bbE}{\mathbb{E}}

\newcommand{\half}{\frac{1}{2}}

\newcommand{\lieg}{\mathfrak{g}}

\newcommand{\lieh}{\mathfrak{h}}

\newcommand{\liem}{\mathfrak{m}}

\newcommand{\RiemN}{R^{N}\relax}
\newcommand{\Riempi}{R^{M}\relax}
\newcommand{\interior}{\iota}

\newcommand{\vece}{\bm{e}}

\theoremstyle{plain}
\newtheorem{thm}{Theorem}[section] 
 
\newtheorem{prop}[thm]{Proposition} 

\theoremstyle{remark}
\newtheorem{comment}{Comment}

\newcommand{\mytitle}{Schwarzschild Spacetime\\ without Coordinates}
\newcommand{\mysubject}{Coordinate Free Schwarzschild}

\begin{document}

\begin{titlepage}

\strut
\par\vspace{1in}

\begin{center}
    {\bf\large \mytitle
    } \\[.5in]
    {\bf Orlando Alvarez
    \footnote{email: 
    \href{mailto:oalvarez@miami.edu?subject=\mysubject}{\texttt{oalvarez@miami.edu}}}}\\[.1in]
    {\em Department of Physics}\\
    {\em University of Miami}\\
    {\em P.O. Box 248046}\\
    {\em Coral Gables, FL 33124 USA}
\end{center}

\vspace*{.3in}
\begin{abstract}
We discuss how to construct the full Schwarzschild (Kruskal-Szekeres)
spacetime in one swoop by using the bundle of orthonormal Lorentz
frames and the Einstein equation without the use of coordinates.  We
never have to write down the Kruskal-Szekeres or an equivalent form of
the metric.

\end{abstract}
\end{titlepage}

\newpage
\small
\tableofcontents
\normalsize
\setcounter{page}{2}

\newpage
\section{Introduction}
\label{sec:intro}

The Schwarzschild solution is probably the most studied nontrivial
solution to the Einstein equations\footnote{For a comprehensive study
of known solutions to the Einstein equations
see~\cite{einstein-equations}.}.  The exterior solution represents
spherically symmetric stars and the interior has a black hole.  What
drives much of the intellectual curiosity of students of general
relativity is the presence of the black hole and its consequences.

I have taught the general relativity course several times over a 20
year period and I have not been completely satisfied with the
discussions of black holes in the introductory course.  The
presentation usually involves deriving the Schwarzschild metric in
standard Schwarzschild coordinates then transforming to
Eddington-Finkelstein coordinates to study what happens as one crosses
the horizon and finally a discussion of the maximal extension in
Kruskal-Szekeres coordinates, see for
example~\cite{carroll:book,dinverno:book,MTW}.  Of course, you could
take as a starting point the Kruskal-Szekeres solution.  This is
neither physically or mathematically satisfying because the radius $r$
of the symmetry $2$-spheres is implicitly given in terms of the
Kruskal-Szekeres $(T,R)$-coordinates by
\begin{equation*}
    T^{2}-R^{2} = \left( 1-\frac{r}{2GM}\right) e^{r/2GM}\,.
\end{equation*}

Another approach is to introduce a lot more mathematical machinery
\cite{Hawking-Ellis:book,Wald:book} and discuss global causal
structures and singularity structures of lorentzian manifolds but this
is overkill if you just want to talk about the Schwarzschild solution.

I wanted to find a middle ground where you could see the whole
extended Schwarzschild solution at once with the geometry and the
physics transparent.  In fact I wanted to find a coordinate
independent way of describing the Schwarzschild solution.  It began by
trying to understand what Birkhoff's Theorem tells you about the
bundle of Lorentz orthonormal frames.  In the process I found such a
coordinate independent geometric approach but unfortunately it is not
elementary at the level of an introductory general relativity course.
It requires much more mathematics, especially an understanding of
group actions on manifolds, of principal fiber bundles~\cite{KN:I} and
of riemannian submersions~\cite{ONeill:submersion}.  It is more than
overkill, nevertheless, I believe the approach to be a new novel and
insightful way of studying the Schwarzschild solution.

This article is not meant to be an exhaustive discussion of the
Schwarzschild spacetime.  I will pick and choose several topics that
are of interest because of the mathematical methods I use.  One
glaring omission is the discussion of the actual singularity at $r=0$.
The reason is that I have no new insight to offer.

In brief, the goal is to derive the extended Kruskal-Szekeres
spacetime without ever writing coordinates.  Instead of studying the
geometry in the Schwarzschild spacetime $N$ directly, we work
``upstairs'' in the bundle of orthonormal Lorentz frames
$\Orthframe(N)$ and indirectly work out the properties of the
spacetime.  By using the Einstein equations and some global structures
in $\Orthframe(N)$ we construct the full spacetime at once.  There is
no ``extension process'' where you begin with the exterior
Schwarzschild solution and find the maximal analytic extension.

We begin by discussing the Cartan structural equation for the bundle 
of orthonormal frames of a semi-riemannian manifold. We begin to 
specialize by studying the restrictions imposed on the structural 
equation if the manifold is a fiber bundle. We do a further 
specialization to the case that the manifold is a semi-riemannian 
submersion. We finally study the case that the semi-riemannian 
submersion arises due to a group action. The Schwarzschild 
spacetime is a manifold of this type. By studying the properties of 
the structural equations we can construct the full Kruskal-Szekeres 
spacetime.

There are extensive computations in these notes because the methods
are not familiar to most physicists or mathematicians.  It makes
extensive use of Cartan's method of the moving frame beyond what most
people use.  It is more of an abstract use of Cartan's method than the
explicit direct computational approach seen in some relativity
textbooks.  There are some nice uses of the machinery.  For this
reason some sections are expository in nature and are not directly
related to the main topic.

\section{Frame Bundles}
\label{sec:frame}

We wish to globally study a semi-riemannian
manifold~\cite{ONeill:SRG}.  Most of the mathematical framework we
will be developing works in both the riemannian case and in the
lorentzian case.  The riemannian language is more standard and for
simplicity I will phrase the discussion as if the manifold was
riemannian.  Of course when we get to black holes we have to work in
the lorentzian framework.  The only times we have to be careful is if
we have a null vector in a subspace.  When we encounter such a case I
will be extra careful.  

We assume that we have an orientable riemannian manifold.  The metric
allows us to consider the orthonormal bases at each tangent space
$T_{x}N$ for $x\in N$.  The collection of all such orthonormal bases
gives us a principal fiber bundle $\Orthframe(N)$, the bundle of all
orthonormal frames.  This bundle has structure group $\SOrth(n)$ where
$n=\dim N$.  Note that $\dim \Orthframe(N) = n +\half n(n-1)$.  One of
the most important properties of $\Orthframe(N)$ is that it has a
canonical global coframing~\cite{KN:I}.  There are $n$ tautologically
defined global $1$-forms\footnote{These are sometimes called the
``soldering forms'' in the older mathematical literature and in some
of the physics literature.} on $\Orthframe(N)$ that will be denoted by
$\{ \theta^{\mu}\}$.  There is the unique Levi-Civita connection on
$\Orthframe(N)$ that gives $\half n(n-1)$ globally defined $1$-forms
$\{\omega^{\mu}{}_{\nu}\}$ with $\omega_{\mu\nu}=-\omega_{\nu\mu}$.
Together the $\half n(n+1)$ $1$-forms
$\{\theta^{\mu},\omega^{\nu}{}_{\rho}\}$ gives a global coframing of
$\Orthframe(N)$.  The dual basis of vector fields is denoted by
$(\vece_{\mu},\vece_{\nu\rho})$.  The important observation is that we
do not have global coordinates on $\Orthframe(N)$ but we have
something that is almost as good, a global coframe. The frame bundle 
is the global structure that we are going to use to study the 
Schwarzschild spacetime.

The Cartan structural equations for the orthonormal frame bundle
$\Orthframe(N)$ of a riemannian manifold $N$ are~\cite{KN:I}
\begin{equation}
    \begin{split}
        d\theta^{\mu} &= -\omega_{\mu\nu}\wedge \theta^{\nu}\,,  \\
	d\omega_{\mu\nu} &= -\omega_{\mu\lambda}\wedge
	\omega_{\lambda\nu} + \half\, R^{N}_{\mu\nu\rho\sigma}
	\theta^{\rho} \wedge \theta^{\sigma}\,.
    \end{split}
    \label{eq:CartanN}
\end{equation}
Note that these are equations on $\Orthframe(N)$ and therefore
$R^{N}_{\mu\nu\rho\sigma}$ are globally defined functions on
$\Orthframe(N)$ with certain equivariance transformation laws under 
the group action.  If you consider a local section $s: U\subset N \to
\Orthframe(N)$ then the pullback $1$-forms $\vartheta^{\mu}
=s^{*}\theta^{\mu}$ give a local orthonormal coframe on $U\subset N$,
the pullback $1$-forms $s^{*}\omega_{\mu\nu} = \gamma_{\mu\nu\rho}
\vartheta^{\rho}$ give the connection coefficients
$\gamma_{\mu\nu\rho}$ in the local orthonormal coframe, and
$s^{*}R^{N}_{\mu\nu\rho\sigma} = R^{N}_{\mu\nu\rho\sigma} \circ s$ are
the components of the Riemann curvature tensor in the local coframe.

Assume we have a local orthonormal coframe $\vartheta^{\mu}$ on
$U\subset N$ with Levi-Civita connection $\Gamma_{\mu\nu}
=\gamma_{\mu\nu\rho}\vartheta^{\rho}$.  Locally we have a
trivialization $U \times \SOrth(n)$ of $\Orthframe(N)$.  If  $(x,g)
\in U \times \SOrth(n)$ then the coframing of $\Orthframe(N)$ may be
locally expressed in terms of the trivialization as
\begin{equation}
    \begin{split}
        \theta^{\mu} &= g^{-1}\vartheta^{\mu}\,, \\
        \omega &= g^{-1}\,dg + g^{-1}\Gamma g\,. \\
    \end{split}
    \label{eq:local-connection}
\end{equation}
If $x^{\mu}$ are local coordinates on $U$ then $(x,g)$ parametrize 
$U\times \SOrth(N)$ and $dx$ and $g^{-1}dg$ are linearly independent 
on the frame bundle. We physicists usually work downstairs and we 
usually think of $\omega_{\mu\nu}$ as $\Gamma_{\mu\nu\rho} 
dx^{\rho}$. Do not do this in this article. The connection lives 
upstairs! Some formulas look different because we are working upstairs.

Next we discuss how to think about the covariant derivative.  A tensor
on the base is viewed upstairs as an ordinary vector valued function
$\xi^{A}$ (a column vector) that has special transformation laws.
Assume we are at a frame\footnote{Think of a frame as a row vector.}
$q = (\vece_{1},\ldots\vece_{n})$ and we act on the frame on the right
by a rotation matrix $g$ where we find that $q'=qg$ then we want
$\xi(qg) = \sigma(g)^{-1}\xi(q)$ where $\sigma$ is a
representation of $\SOrth(n)$.  The differential of $\xi$ is a
$1$-form and we have to specify how $\xi$ changes both along the
fiber and transverse to the fiber.  We know that $d\xi$ must be
expressible as a linear combination of the coframe
$(\theta^{\mu},\omega_{\nu\rho})$.  The question is which linear
combination.  Part of the definition of a connection~\cite{KN:I} is
that $\omega$ restricted to the vertical tangent space is the left
invariant form on the Lie algebra.  In local formulas
\eqref{eq:local-connection} we see that $\omega = g^{-1}dg$ when
tangent to the fibration (given by $dx=0$). We know how $\xi$ 
transforms under the $\SOrth(n)$ action and thus we conclude that
\begin{equation}
    d\xi = -\dot{\sigma}(\omega)\xi + 
    \xi_{;\mu}\theta^{\mu}\,.
    \label{eq:cov-der}
\end{equation}
Here $\dot{\sigma}$ is the induced Lie algebra representation.  The
``horizontal'' component of $d\xi$ is denoted by
$\xi_{;\mu}\theta^{\mu}$ and it is called the \emph{covariant
differential}.

If a vector field 
\begin{equation}
    V = V^{\mu} \vece_{\mu} + \half V^{\mu\nu} \vece_{\mu\nu}
    \label{eq:V-frame}
\end{equation}
on $\Orthframe(N)$ generates an isometry then 
$\Lieder_{V}\theta^{\nu}=0$. A simple computation shows 
that\footnote{We use the traditional semi-colon notation to denote 
covariant derivatives.}
\begin{equation*}
    0 = -V_{\mu\nu} \theta^{\nu} + \omega_{\mu\nu} V^{\nu}  + 
    dV_{\mu} = -V_{\mu\nu}\theta^{\nu} + V_{\mu;\nu}\theta^{\nu}\,.
\end{equation*}
We immediately learn two things
\begin{equation}
    \begin{split}
    0 & = V_{\mu;\nu} + V_{\nu;\mu}  \,,\\
    V_{\mu\nu} &= \half( V_{\mu;\nu} - V_{\nu;\mu})\,.
    \end{split}
    \label{eq:Killing}
\end{equation}
The first of the above are Killing's equations.  The second one
determines the $\vece_{\mu\nu}$ component of the vector field $V$, see
\eqref{eq:V-frame}.  Next we observe that $0 = d \Lieder_{V}
\theta^{\mu} = \Lieder_{V}(d\theta^{\mu})$.  Using the Cartan
structural equation we immediately see that
$(\Lieder_{V}\omega_{\mu\nu}) \wedge \theta^{\nu} = 0$.  An
application of the corollary to Cartan's lemma, see
Appendix~\ref{sec:cartan-lemma}, yields that $\Lieder_{V}
\omega_{\mu\nu} = 0$, \emph{i.e.}, the connection is invariant under
the infinitesimal isometry.

We will be copiously using the differential forms version of the
Frobenius Theorem~\cite{Warner}.  To avoid too much terminology we
state the theorem in the following practical way.
\begin{thm}[Frobenius]
    \label{thm:frobenius}
    Assume that on a manifold $M$ we have a collection of $k$ linearly
    independent non-vanishing $1$-forms $\{\varphi^{\alpha}\}$ with
    the property that $d\varphi^{\alpha} =
    \xi^{\alpha}{}_{\beta}\wedge \varphi^{\beta}$ for some $1$-forms
    $\xi^{\alpha}{}_{\beta}$.  Then through every $q\in M$ there
    exists a unique maximal connected submanifold $S_{q}$ containing
    $q$ with $\dim S_{q} = \dim M - k$ such that for every vector
    field $X$ tangent to $S_{q}$ we have that $\varphi^{\alpha}(X)=0$.
\end{thm}
The integrability conditions on the differential forms are sometimes
written as ``$d\varphi^{\alpha} = 0 \mod \varphi$''.

\section{Fibration}
\label{sec:fibration}

Let $\pi:N\to M$ be a fiber bundle where $N$ is a riemannian manifold.
The fibers are isomorphic to a manifold $F$.  If $x\in M$ then the
fiber over $x$ will be denoted by $F_{x}$.  Vectors tangent to the
fiber will be called vertical and vectors orthogonal to the fiber will
be called horizontal.  If $\dim N =n$ and $\dim M =p$ then the
dimension of the fibers is $q=n-p$.  The existence of the fibration
allows for a reduction of the structure group $\SOrth(n)$ of
$\Orthframe(N)$ to $\SOrth(p)\times \SOrth(q)$ obtaining a principal
sub-bundle $\Redframe(N) \subset \Orthframe(N)$.  If we introduce
indices $i,j,k,\dots$ ``associated to M'' to run from $1,\dots,p$ and
indices $a,b,c,d,\dots$ ``associated to the fibers'' to run from
$p+1,\dots,n$ then the first structural equation may be written as
\begin{equation}
    \begin{split}
        d\theta^{i} &= - \omega_{ij}\wedge \theta^{j} - \omega_{ia} 
        \wedge \theta^{a}\,,	\\
        d\theta^{a} &= -\omega_{ab}\wedge \theta^{b} - \omega_{ai} 
        \wedge \theta^{i}\,.
    \end{split}
    \label{eq:Cartan1}
\end{equation}
Once the structure group is reduced we have that the $\omega_{ai}$ 
become torsion:
\begin{equation}
    \omega_{ai} = K_{abi}\theta^{b} - M_{ija}\theta^{j}\,.
    \label{eq:torsion}
\end{equation}
This requires some explanation and is best understood by looking at
local expression \eqref{eq:local-connection}.  Once we are on
$\Redframe(N) \subset \Orthframe(N)$ we can no longer move along the
group directions that are not tangent to $\Redframe(N)$.  This means
that $g^{-1}dg$ restricted to $\Redframe$ vanishes in Lie algebra
directions orthogonal to $\so(p) \oplus \so(q) \subset \so(n)$.  Thus
when restricted to $\Redframe(N)$ we only get the $\Gamma$ part of
$\omega$ in \eqref{eq:local-connection}.  Schematically we have that
$\Gamma = \gamma \vartheta = \gamma g \theta$ and this is how
\eqref{eq:torsion} arises.  With this in mind we see that the first of
\eqref{eq:Cartan1} becomes
\begin{equation}
    d\theta^{i} = -\omega_{ij}\wedge\theta^{j} -M_{ija} \theta^{j} 
    \wedge \theta^{a} - \half(K_{abi}-K_{bai}) \theta^{a} \wedge 
    \theta^{b}\,.
    \label{eq:c1}
\end{equation}
The pfaffian equations $\theta^{i}=0$ determine an integrable vertical 
distribution that defines the fibration and therefore the Frobenius theorem 
requires
\begin{equation}
    K_{abi}=K_{bai}\,.
    \label{eq:second}
\end{equation}
This is the statement that the second fundamental form for the 
submanifolds associated with the fibration is symmetric. It is 
worthwhile to consider the symmetric and anti-symmetric parts of $M$:
\begin{equation}
    \begin{split}
    S_{ija} & = \half(M_{ija} + M_{jia})\,, \\
    A_{ija} & = \half(M_{ija} - M_{jia})\,.
    \end{split}
    \label{eq:defSA}
\end{equation}
The structure equation may be written as
\begin{equation*}
    d\theta^{i} = -\omega_{ij}\wedge\theta^{j} 
    +A_{ija}  \theta^{a}\wedge \theta^{j} 
    -S_{ija} \theta^{j} \wedge \theta^{a} \,.
\end{equation*}
Following Cartan we try to absorb as much torsion as possible by 
defining a new connection
\begin{equation}
    \pi_{ij} = \omega_{ij} -A_{ija}\theta^{a}\,.
    \label{eq:defpiconn}
\end{equation}
The structural equation becomes
\begin{equation*}
    d\theta^{i} = -\pi_{ij}\wedge\theta^{j} 
    -S_{ija} \theta^{j} \wedge \theta^{a} \,.
\end{equation*}

Similarly, the second of \eqref{eq:Cartan1} becomes
\begin{equation}
    d\theta^{a} = -\omega_{ab}\wedge\theta^{b} 
    -K_{abi}\theta^{b}\wedge\theta^{i}
    - A_{ija} \theta^{i}\wedge\theta^{j}\,.
    \label{eq:c2}
\end{equation}
The vanishing of the tensor $A_{ija}$ is the integrability
condition for the distribution defined by $\theta^{a}=0$.
You cannot absorb the torsion in $d\theta^{a}$ because $K_{abi}$ is 
symmetric under $a \leftrightarrow b$.

Finally we make the following useful remark. If $X$ is a horizontal 
vector field ,\emph{i.e.}, $\interior_{X}\theta^{a}=0$ then
\begin{equation}
    \Lieder_{X}(\theta^{a} \otimes \theta^{a}) = 2X^{i} K_{abi} \,
    \theta^{a}\otimes \theta^{b} - 2 X^{i}
    A_{ija}(\theta^{j}\otimes\theta^{a} +
    \theta^{a}\otimes\theta^{j})\,.
    \label{eq:Lie-der-K}
\end{equation}

If the horizontal distribution is integrable then $A_{ija}=0$ and the 
equation above simplifies to
\begin{equation}
    \Lieder_{X}(\theta^{a} \otimes \theta^{a}) = 2X^{i} K_{abi} \,
    \theta^{a}\otimes \theta^{b} \,.
    \label{eq:Lie-der-K-1}
\end{equation}
If $\eta_{F}$ is the volume element on the fiber then
\begin{equation}
    \Lieder_{X} \eta_{F} = X^{i} K^{a}{}_{ai}\, \eta_{F}\,.
    \label{eq:Lie-der-K-2}
\end{equation}

\textbf{SUMMARY:} The structural equations for a fibration are
\begin{equation}
    \begin{split}
        K_{abi} &= K_{bai}\,,\\
	M_{ija} &= S_{ija} + A_{ija}\,,\text{ see \eqref{eq:defSA}},  \\
        \omega_{ai} &= K_{abi}\theta^{b} - M_{ija}\theta^{j}\,,  \\
	\pi_{ij} &= \omega_{ij} -A_{ija}\theta^{a}\,, \\
        d\theta^{i} &= -\pi_{ij}\wedge\theta^{j} 
        -S_{ija}\theta^{j}\wedge\theta^{a}\,,  \\
        d\theta^{a} &= -\omega_{ab}\wedge\theta^{b} 
        -K_{abi}\theta^{b}\wedge\theta^{i} - 
        A_{ija} \theta^{i}\wedge\theta^{j}\,.
    \end{split}
    \label{eq:fibration}
\end{equation}

\subsection{Local Description}
\label{sec:local}

Assume our total space $N$ is euclidean space $\bbE^{n}$ and that
after a rotation we locally describe the fibers near the origin as the
level sets of the $p$ functions
\begin{equation}
    f^{i}(x) = x^{i} + \half h_{abi}\,x^{a}x^{b} + O(x^{3})\,,
    \label{eq:level-sets}
\end{equation}
where we use cartesian coordinates $(x^{i},x^{a})$.  We know that
$D^{N}_{e_{a}}e_{i} = e_{b}K_{abi} + e_{j} \omega^{j}{}_{i}(e_{a})$.
Taking the gradient of the function that defines the level sets we see
that the normals (to leading order) are given by
\begin{equation*}
    e_{i} = \frac{\partial}{\partial x^{i}} + h_{abi}x^{b}\; 
    \frac{\partial}{\partial x^{a}}\,.
\end{equation*}
Comparing with the connection definition of the extrinsic curvatures
we see that at the origin we have that $K_{abi} = h_{abi}$.  Notice
that $K_{abi}$ is a first order invariant of the metric, \emph{i.e.},
$K \sim \partial g$, while the curvature is second order $R \sim
\partial^{2}g$.

For the special case $K_{abi} = \delta_{ab}S_{i}$ we see that locally 
our embedded $p$-surface looks like a surface of revolution about the 
normal $S$ direction.
\subsection{Totally Geodesic Fibration}
\label{sec:totally-geodesic}

Let's forget the frame bundle and work on the base manifold $N$.  The
extrinsic curvature tensor (second fundamental forms) is defined by
$K(V,W,X) = (V,D^{N}_{W}X)$ where $V$, $W$ are vertical vectors and
$X$ is a horizontal vector.  In a local frame we have $K_{abi} =
(e_{a},D^{N}_{e_{b}}e_{i})$.  A geodesic with tangent vector field $V$
that is tangent to a fiber will satisfy the geodesic equation
$D^{N}_{V}V=0$ and tangentiality condition $(V,X)=0$ for all
horizontal vectors $X$.  Since the connection $D^{N}$ is metric
compatible we have $0 = D^{N}_{V}(V,X) = (D^{N}_{V}V,X) +
(V,D^{N}_{V}X) = K(V,V,X)$ for all possible $V$.  This implies that
the second fundamental form must vanish if all geodesics are tangent 
to the fibration.

\subsection{Gauss and Codazzi Equations}
\label{sec:gauss-codazzi}

The integrability of the vertical distribution allows us to 
consistently substitute $\theta^{i}=0$ into the equations above by 
restricting to $F_{x}$, the fiber over $x\in M$.
\begin{equation*}
    \begin{split}
        K_{abi} &= K_{bai}\,,\\
	M_{ija} &= S_{ija} + A_{ija}\,,\text{ see \eqref{eq:defSA}},  \\
        \omega_{ai} &= K_{abi}\theta^{b}\,,  \\
	\pi_{ij} &= \omega_{ij} -A_{ija}\theta^{a}\,, \\
        d\theta^{a} &= -\omega_{ab}\wedge\theta^{b}\,.
    \end{split}
\end{equation*}
Note that
\begin{equation*}
    d\theta^{a}\rvert_{F_{x}} = -\omega_{ab}\wedge 
    \theta^{b}\rvert_{F_{x}}
\end{equation*}
which tells us that $\omega_{ab}\rvert_{F_{x}}$ is the torsion free
riemannian connection on $F_{x}$.  To work out the curvature we
observe that
\begin{equation*}
    \half R^{N}_{ab\mu\nu} \theta^{\mu}\wedge\theta^{\nu} = 
    d\omega_{ab} + \omega_{ac}\wedge\omega_{cb} - 
    \omega_{ai}\wedge\omega_{bi}\,.
\end{equation*}
Restricting to $F_{x}$ we get the Gauss equation
\begin{equation}
	\half R^{N}_{abcd} \theta^{c}\wedge\theta^{d} = 
	\half R^{F_{x}}_{abcd} \theta^{c}\wedge\theta^{d}
	- \half(K_{aci}K_{bdi} -K_{adi}K_{bci})
	\theta^{c}\wedge\theta^{d}\,. 
    \label{eq:curvfiber}
\end{equation}
This is often written
\begin{equation}
    R^{N}_{abcd} = R^{F_{x}}_{abcd} 
    - (K_{aci}K_{bdi}-K_{adi}K_{bci})\,.
    \label{eq:gauss}
\end{equation}

Next we derive the Codazzi equation. Note that 
$\omega_{ai} = K_{abi}\theta^{b}$ when restricted to $F_{x}$. We have 
\begin{equation*}
    \begin{split}
    \half R^{N}_{aicd}\theta^{c}\wedge \theta^{d} &=
    \half R^{N}_{ai\mu\nu}\theta^{\mu}\wedge \theta^{\mu}\rvert_{F_{x}} \,\\
    & =
    \left( d\omega_{ai} + \omega_{ab}\wedge\omega_{bi} + 
    \omega_{aj}\wedge\omega_{ji} \right) \rvert_{F_{x}}\,,\\
    &  = 
    \left( d\omega_{ai} + \omega_{ab}\wedge\omega_{bi} + 
    \omega_{aj}\wedge\pi_{ji} \right) \rvert_{F_{x}}
    + \omega_{aj}\wedge A_{jid}\theta^{d}\,,\\
    &=
    D(K_{adi}\theta^{d})\rvert_{F_{x}} + K_{acj}A_{jid} 
    \theta^{c}\wedge\theta^{d}\,.
    \end{split}
\end{equation*}
In the above $D$ is the covariant differential with connection 
$(\omega_{ab},\pi_{ij})$. If we write $DK_{abi} = 
K_{abi;j}\theta^{j} + K_{abi;c}\theta^{c}$ then the last line of the 
above may be written as
\begin{equation*}
    K_{adi;c}\theta^{c}\wedge\theta^{d}+ K_{acj}A_{jid} 
    \theta^{c}\wedge\theta^{d}\,.
\end{equation*}
We have derived the Codazzi equation
\begin{equation}
    R^{N}_{aicd} = (K_{adi;c}-K_{aci;d}) + 
    (K_{acj}A_{jid}-K_{adj}A_{jic})\,.
    \label{eq:codazzi}
\end{equation}

\section{Riemannian Submersion}
\label{sec:submersion}

Many of the spacetimes studied by physicists are semi-riemannian
submersions.  A submersion $\pi:N\to M$ of riemannian manifolds is
called a \emph{riemannian submersion} if $d\pi$ preserves the inner
product of vectors orthogonal to the fibers~\cite{ONeill:submersion}.
A tangent vector is \emph{horizontal} if it is orthogonal to the
fibers.  A riemannian submersion implies a very specific form for the
metric.  If $x^{i}$ are local coordinates on the base $M$ and if
$y^{a}$ are local coordinates on the fiber $F$ then $(x,y)$ are local
coordinates on $N$.  The fibers are the submanifolds with $x$ fixed.
The metric of a submersion is locally of the form
\begin{equation*}
    ds^{2}_{N} = g_{ij}(x) dx^{i}\, dx^{i} + g_{ab}(x,y) \left(dy^{a} + 
    C^{a}{}_{i}(x,y)dx^{i}\right) \left(dy^{b} + 
    C^{b}{}_{j}(x,y)dx^{j} \right)\,.
\end{equation*}
Sometimes in the physics literature this is referred to as a metric of
Kaluza-Klein type.  If we fix $x$ then the metric on a fiber
$g_{ab}(x,y) dy^{a}\,dy^{b}$ varies as we move along the base.  In
general curves of constant $x$ will not be orthogonal to curves of
constant $y$.  On the other hand we have that $\partial/\partial
y^{a}$ is orthogonal to the horizonal vector field $\partial/\partial
x^{i} - C^{a}{}_{i}\, \partial/\partial y^{a}$.  The metric on the
horizontal space is $g_{ij}(x)$ which is the metric on the base and is
independent of choice of $y$.  O'Neill studied the properties of
riemannian submersions and discovered that the geometry was governed
by two tensor fields.  One tensor field is the second fundamental form
(the extrinsic curvature) of the embedding of the fibers in $N$ and
the other tensor field is the integrability tensor for the horizontal
spaces.  We present here a formulation that is equivalent to O'Neill's
except that everything is expressed in terms of orthonormal frames
adapted to the fibration.

We first do some local analysis.  Locally pull back the $\theta$s from
$\Redframe(N)$ to $N$ via a section.  If $V$ is a vertical vector
field on $N$, \emph{i.e.}, tangential to the fibration $\pi:N\to M$,
the condition of a riemannian submersion can be written as
$\Lieder_{V}(\theta^{i}\otimes \theta^{i}) = 0$.  This is simply the
statement that $\theta^{i} \otimes \theta^{i}$ descends to the base.
A vertical vector field satisfies $\iota_{V}\theta^{i} =0$.  A simple
computation shows that
\begin{equation*}
    \begin{split}
    \Lieder_{V}(\theta^{i}\otimes \theta^{i}) &= - (\pi_{ij}(V) + 
    \pi_{ji}(V)) \theta^{i}\otimes\theta^{j}
    +2 V^{a}S_{ija} \theta^{i}\otimes\theta^{j}\,,\\
    &= 2 V^{a}S_{ija} \theta^{i}\otimes\theta^{j}\,.
    \end{split}
\end{equation*}
The degenerate quadratic form $\theta^{i} \otimes \theta^{i}$ on $N$
descends to a positive definite quadratic form on the base $M$ if
\begin{equation}
    S_{ija} = 0\,.
    \label{eq:Lskew}
\end{equation}

We can do the same analysis on $\Redframe(N)$  but the equations look 
different. The vector field in this case is a vector field on 
$\Redframe(N)$ and will be of the form
\begin{equation*}
    V = V^{a} \vece_{a} +\half V^{ab} \vece_{ab} + \half  V^{ij} 
    \vece_{ij} \,.
\end{equation*}
Here $(\vece_{a},\vece_{i},\vece_{ab},\vece_{ij})$ is the basis dual
to $(\theta^{a},\theta^{i},\omega_{ab},\pi_{ij})$.  We want
$\Lieder_{V}\theta^{i} = 0$ and a brief computation leads to the
equation $0 = -V_{ij}\theta^{j} + S_{ijb}V^{b}\theta^{j}$.  From this
we learn that $V_{ij}=0$ and $S_{ija}=0$.  Note that the vector field
\begin{equation*}
    V = V^{a} \vece_{a} +\half V^{ab} \vece_{ab}  \,.
\end{equation*}
is of the type that is associated intrinsically with the fibers of the
fibration $\pi:N \to M$.  Finally we note that the conditions arising
from $\Lieder_{V}\theta^{i}=0$ at $z\in N$ do not depend on the
derivatives of the components of $V$ and therefore depend only on
$V(z)$ and not its extension to a neighborhood of $z$.

The structural equations for a riemannian submersion are
\begin{equation}
    \begin{split}
	K_{abi} &= K_{bai}\quad\text{and}\quad A_{ija} = - A_{jia}\,,\\
	\omega_{ai} &= K_{abi}\theta^{b} - A_{ija}\theta^{j}\,,  \\
	\pi_{ij} &= \omega_{ij} -A_{ija}\theta^{a}\,, \\
	d\theta^{i} &= -\pi_{ij}\wedge\theta^{j}\,,  \\
	d\theta^{a} &= -\omega_{ab}\wedge\theta^{b} 
	-K_{abi}\theta^{b}\wedge\theta^{i} - 
	A_{ija} \theta^{i}\wedge\theta^{j}\,.
    \end{split}
    \label{eq:submersion}
\end{equation}

As we mentioned before $K_{abi}$ are the second fundamental forms 
(extrinsic curvatures) of 
the fibration. The tensor $A_{ija}$ measures the integrability of the 
horizontal tangent spaces. We use the basic relation that if $\xi$ is 
a $1$-form then $d\xi(X,Y) = X(\xi(Y)) -Y(\xi(X))-\xi([X,Y])$. First 
we take a section and we pullback the structural equation. We 
have that $\vece_{i}$ is a basis for the horizontal spaces of the 
submersion. If the horizontal spaces are to be integrable then the 
bracket of two horizontal vector fields must be horizontal. We 
compute the vertical  component of the bracket as follows:
\begin{equation*}
    \theta^{a}([\vece_{i},\vece_{j}]) =
    \vece_{i}(\theta^{a}(\vece_{j})) +
    \vece_{j}(\theta^{a}(\vece_{i})) -
    (d\theta^{a})(\vece_{i},\vece_{j}) = 2 A_{ija}\,.
\end{equation*}
Thus we see that the horizontal distribution is integrable if and 
only if the integrability tensor $A_{ija}$ vanishes.

We note that $d^{2}\theta^{i}=0$ and therefore
\begin{equation}
    \Pi_{ij}\wedge\theta^{j}=0\,,
    \label{eq:dpi}
\end{equation}
where
\begin{equation}
    \Pi_{ij} = d\pi_{ij} + \pi_{ik}\wedge\pi_{kj}\,.
    \label{eq:defpi}
\end{equation}
Wedging \eqref{eq:dpi} with $\theta^{k_{1}}\wedge\dots\wedge
\theta^{k_{p-1}}$ we conclude that $\Pi_{ij} \equiv 0 \mod
\theta^{k}$.  So we can write $\Pi_{ij} = \Psi_{ijk}\wedge\theta^{k}$
for some $1$-forms $\Psi_{ijk}$ that are skew in the indices $i
\leftrightarrow j$.  From \eqref{eq:dpi} we see that 
$\Psi_{ijk}\wedge \theta^{j}\wedge \theta^{k} = 0$. This tells us 
that $(\Psi_{ijk}-\Psi_{ikj}) \equiv 0 \mod \theta^{l}$. Since 
$\Psi_{ijk}$ is symmetric in $j \leftrightarrow k$ modulo 
$\theta^{l}$ but is it skew in $i \leftrightarrow j$ we conclude 
that $\Psi_{ijk} \equiv 0 \mod \theta^{l}$. This tells us that 
$\Psi_{ijk} = P_{ijkl}\theta^{l}$. Putting this all together we conclude 
that
\begin{equation}
    \Pi_{ij} = d\pi_{ij} + \pi_{ik}\wedge\pi_{kj} =
    \half R^{M}_{ijkl}\theta^{k}\wedge\theta^{l}\,.
    \label{eq:curvonM}
\end{equation}
Note that the right hand side is horizontal!  This would not be true
if we had used the $\omega_{ij}$ connection.

In the same way we see that the structural equation for $\omega_{ab}$ 
is given by
\begin{equation}
    \begin{split}
	d\omega_{ab} &= -\omega_{ac}\wedge \omega_{cb} +\half 
	R^{F_{x}}{}_{abcd} \theta^{c}\wedge \theta^{d} \\
	&\quad + \left( R^{N}{}_{abcj} -A_{ijb}K_{aci} +A_{ija}K_{bci} 
	\right) \theta^{c}\wedge \theta^{j} \\
	&\quad + \half \left( R^{N}{}_{abjk} + A_{ija}A_{ikb} 
	-A_{ika}A_{ijb}\right) 
	\theta^{j}\wedge\theta^{k}
    \end{split}
    \label{eq:cartan-vert}
\end{equation}
where $R^{F_{x}}$ is given by the Gauss equation \eqref{eq:gauss}.

Next we compute the riemannian curvature of the base $M$ by using the 
riemannian data on the bundle $N$. Note that
\begin{equation*}
    \half R^{N}{}_{ij\mu\nu} \theta^{\mu}\wedge \theta^{\nu}
    =d\omega_{ij} +\omega_{ik}\wedge\omega_{kj} - \omega_{ai} \wedge
    \omega_{aj}\,.
\end{equation*}
Substituting the appropriate expressions we find
\begin{equation}
    \tensor{\RiemN}{\down{i}\down{j}\down{k}\down{l}} =
    \tensor{\Riempi}{\down{i}\down{j}\down{k}\down{l}} +
    \tensor{A}{\down{i}\down{l}\down{a}}
    \tensor{A}{\down{j}\down{k}\up{a}} -
    \tensor{A}{\down{i}\down{k}\down{a}}
    \tensor{A}{\down{j}\down{l}\up{a}} - 2
    \tensor{A}{\down{i}\down{j}\down{a}}
    \tensor{A}{\down{k}\down{l}\up{a}}\,.
    \label{eq:riembase}
\end{equation}

\begin{table}[tbp]
    \centering
    \setlength{\extrarowheight}{1pt}
    \begin{tabular}{ccl}
	\toprule
	plane & $\{n\}$ & Curvature \\
	\midrule 
	$ab$ & $\{0\}$ & $R^{N}{}_{abcd} = R^{F_{x}}_{abcd} -
	(K_{aci}K_{bd}{}^{i}-K_{adi}K_{bc}{}^{i})$ \hfill\text{\small 
	[Gauss eq.]}\\
	$ai$ or B & $\{1\}$ & $R^{N}{}_{aibc} = -K_{abi;c} + K_{aci;b} 
	-A_{kib}K_{ac}{}^{k} + A_{kic}K_{ab}{}^{k}$ 
	\hfill\text{\small [Codazzi eq.]}\\
	$ai$ & $\{2\}$ & $R^{N}{}_{aibj} = A_{ikb}A_{j}{}^{k}{}_{a} -
	A_{ija;b} -K_{abi;j} - K_{aci}K_{b}{}^{c}{}_{j}$ \\
	$ai$ & $\{3\}$ & $R^{N}{}_{aijk} = A_{ija;k} - A_{ika;j}
	-2A_{jkb}K_{a}{}^{b}{}_{i}$ \\
	$ij$ & $\{2\}$ & $R^{N}{}_{ijab} = A_{ikb}A_{j}{}^{k}{}_{a}
	-A_{ika}A_{j}{}^{k}{}_{b} -A_{ija;b} +A_{ijb;a} +K_{acj} 
	K_{b}{}^{c}{}_{i} -K_{aci}K_{b}{}^{c}{}_{j}$ \\
	$ij$ & $\{3\}$ & $R^{N}{}_{ijkb} = A_{ijb;k} - 
	A_{jka}K^{a}{}_{bi} + A_{ika}K^{a}{}_{bj} + 
	A_{ija}K^{a}{}_{bk}$ \hfill\text{\small 
	[dual Codazzi eq.]}\\
	$ij$ & $\{4\}$ & $R^{N}{}_{ijkl} = \Riempi_{ijkl} + A_{ila}A_{jk}{}^{a}
	-A_{ika}A_{jl}{}^{a} -2 A_{ija}A_{kl}{}^{a}$ \hfill\text{\small 
	[dual Gauss eq.]}
	\\ \addlinespace
	B & $\{2\}$ & 
	$\displaystyle 
	     \begin{aligned}
		 R^{N}{}_{abjk} &= -A_{ija} A^{i}{}_{kb} +
		 A_{ijb}A^{i}{}_{ka} -2A_{jka;b} \\
		 &\quad -K_{abj;k} +K_{abk;j}
		 -K_{bck} K^{c}{}_{aj} + K_{bcj} K^{c}{}_{ak}
	    \end{aligned}
	$ \\ \addlinespace
	B & $\{3\}$ &
	$\displaystyle
	    \begin{aligned}
		0 &= -A_{jka;l}  - A_{kla;j} - A_{lja;k} \\
		&\quad + A_{jkb}K_{a}{}^{b}{}_{l}
		+ A_{klb}K_{a}{}^{b}{}_{j} 
		+ A_{ljb}K_{a}{}^{b}{}_{k} 
	    \end{aligned}
	$ \\
	\bottomrule
    \end{tabular}
    \begin{quote}
    \caption[xyzfoo]{\small Relationship of the bundle curvature
    to the base geometry and fiber geometry for a riemannian
    submersion.  Equations associated with a rotation in the
    $\mu\nu$-plane, \emph{i.e.}, a consequence of $d\omega_{\mu\nu}$,
    are labeled by the first column.  The second column uses O'Neill's
    notation where $\{n\}$ denotes the number of horizontal indices.
    Rows identified with a ``B'' are equations that are a direct
    consequence of the Bianchi identities that follow from
    $d^{2}\theta^{a}=0$.}
    \label{tbl:curv-submersion}
    \end{quote}
\end{table}

We explicitly state the submersion curvature relations in
Table~\ref{tbl:curv-submersion} using the O'Neill
notation~\cite{ONeill:SRG} where $\{n\}$ denotes the number of base
indices.  If you take $ai$--$\{3\}$ and use the Bianchi identities
B--$\{3\}$ you get $ij$--$\{3\}$.  The term B--$\{2\}$ is not skew
under $a \leftrightarrow b$.  From the fact that $R^{N}{}_{abjk}+
R^{N}{}_{bajk}=0$ we learn that
\begin{equation}
    A_{jka;b}+A_{jkb;a} = -K_{abj;k} + K_{abk;j}\,.
    \label{eq:A-K-der}
\end{equation}
This relationship is also necessary to ensure that $R^{N}{}_{aibj} =
R^{N}{}_{bjai}$.  If you insert this relationship into B--$\{2\}$ you
obtain $ij$--$\{2\}$.  Finally we observe that using the above
relationship we can write a manifestly symmetric expression for
$R^{N}{}_{aibj}$.  We note that by \eqref{eq:A-K-der}, the term
between parentheses in the expression $R^{N}{}_{aibj} =
A_{ikb}A_{j}{}^{k}{}_{a} - K_{aci}K_{b}{}^{c}{}_{j} -( A_{ija;b}
+K_{abi;j})$ is symmetric under the interchange $ai \leftrightarrow
bj$.  Therefore we have
\begin{equation}
    R^{N}{}_{aibj} = A_{ikb}A_{j}{}^{k}{}_{a} -
    K_{aci}K_{b}{}^{c}{}_{j} - \half( A_{ija;b} + A_{jib;a} +K_{abi;j}
    +K_{baj;i})\,.
    \label{eq:raibj-sym}
\end{equation}

Using these relations the structural equation \eqref{eq:cartan-vert}
may be rewritten as
\begin{equation}
    \begin{split}
	d\omega_{ab} &= -\omega_{ac}\wedge \omega_{cb} +\half 
	R^{F_{x}}{}_{abcd} \theta^{c}\wedge \theta^{d} \\
	&\quad + \left( -K_{caj;b} + K_{cbj;a} \right)
	\theta^{c}\wedge \theta^{j} \\
	&\quad + \half \left( -A_{jka;b} + A_{jkb;a} -K_{acj}
	K_{b}{}^{c}{}_{k} + K_{bcj} K_{a}{}^{c}{}_{k}\right)
	\theta^{j}\wedge\theta^{k}
    \end{split}
    \label{eq:cartan-vert-1}
\end{equation}
If we now use the structure equation above look for Bianchi 
identities in $d^{2}\theta^{a}=0$ we find \eqref{eq:A-K-der} and 
B--$\{3\}$ as the identities in addition to the cyclic identity that 
$R^{F_{x}}$ satisfies.

\textbf{SUMMARY:} The full structural equations for a riemannian
submersion are
\begin{equation}
    \begin{split}
	K_{abi} &= K_{bai}\quad\text{and}\quad A_{ija} = - A_{jia}\,,\\
	\omega_{ai} &= K_{abi}\theta^{b} - A_{ija}\theta^{j}\,,  \\
	\pi_{ij} &= \omega_{ij} -A_{ija}\theta^{a}\,, \\
	d\theta^{i} &= -\pi_{ij}\wedge\theta^{j}\,,  \\
	d\theta^{a} &= -\omega_{ab}\wedge\theta^{b} 
	-K_{abi}\theta^{b}\wedge\theta^{i} - 
	A_{ija} \theta^{i}\wedge\theta^{j}\,,\\
	d\pi_{ij} &= -\pi_{ik}\wedge\pi_{kj} + \half R^{M}{}_{ijkl}\, 
	\theta^{k}\wedge\theta^{l}\,, \\
	d\omega_{ab} &= -\omega_{ac}\wedge \omega_{cb} +\half 
	R^{F_{x}}{}_{abcd} \theta^{c}\wedge \theta^{d} \\
	&\quad + \left( -K_{caj;b} + K_{cbj;a} \right)
	\theta^{c}\wedge \theta^{j} \\
	&\quad + \half \left( -A_{jka;b} + A_{jkb;a} -K_{acj}
	K_{b}{}^{c}{}_{k} + K_{bcj} K_{a}{}^{c}{}_{k}\right)
	\theta^{j}\wedge\theta^{k}
    \end{split}
    \label{eq:submersion-full}
\end{equation}

If $X,Y$ are horizontal vectors then the sectional curvature is given 
by
\begin{equation}
    \text{sect}^{N}(X,Y) =\text{sect}^{M}(X,Y) -3\,\frac{(A_{ika}X^{i}Y^{k})
    (A_{jl}{}^{a}X^{j}Y^{l})}{(X,X)(Y,Y)-(X,Y)^{2}}\,.
    \label{eq:sectcurv}
\end{equation}
A consequence of the above is O'Neill's
result~\cite{ONeill:submersion} that in a strictly riemannian
submersion the sectional curvature of the base is ``increased''
because the second summand subtracts a manifestly positive
semi-definite expression.  Note that if $N$ is flat then the base
always has positive sectional curvature.

\subsection{The Ricci Tensor and Ricci Scalar}
\label{sec:ricci-submersion}

It is straightforward to write down the Ricci tensor in the case of a 
submersion:
\begin{equation}
    \begin{split}
    R^{N}{}_{bd} & = R^{F_{x}}{}_{bd} - K_{bd}{}^{i}{}_{;i} - 
    K^{c}{}_{ci}K_{bd}{}^{i} + A^{ik}{}_{b} A_{ikd}\,,
    \\
    R^{N}{}_{ai} & = -K^{b}{}_{bi;a} +K^{b}{}_{ai;b} + A^{j}{}_{ia;j}
    -A_{ijb}K_{ab}{}^{j} + A_{kia}K_{b}{}^{bk}\,,
    \  \\
    R^{N}{}_{ij} & = R^{M}{}_{ij} -2 A_{ikd}A_{j}{}^{kd} - 
    K_{cdi}K^{cd}{}_{j} - \half( K^{a}{}_{ai;j} + K^{a}{}_{aj;i})\,.
    \end{split}
    \label{eq:ricci-submersion}
\end{equation}
The Ricci scalar is easily seen to be
\begin{equation}
    R^{N} = R^{M} + R^{F_{x}} - 2 K_{a}{}^{ai}{}_{;i} -K_{abi}K^{abi} 
    -K^{c}{}_{ci}K_{a}{}^{ai} -A_{ija}A^{ija}\,.
    \label{eq:ricci-scalar}
\end{equation}

\section{The Group Action}
\label{sec:group}

\subsection{Transitive Case and Invariant Tensors}
\label{sec:group-transitive}

Assume we have a connected Lie group $G$ acting transitively on a
connected manifold $M$ via isometries.  Let $H \subset G$ be the
isotropy group at $x\in M$.  We know that $M \approx G/H$ and that
$\pi: G \to M$ is a principal $H$-bundle.  We assume that the Lie
algebra is a reductive Lie algebra: $\lieg = \liem \oplus \lieh$ with
$[\lieh, \liem] \subset \liem$.  There is a canonical identification
of $T_{x}M$ with $\liem$.  It is well known that $G$-invariant tensors
on $M \approx G/H$ are in a 1-1 correspondence with $H$-invariant
tensors on $\liem$, see \cite{ONeill:SRG}.  The argument is roughly as
follows.  Assume $S$ is a $G$-invariant tensor on $M$ then if $y = g
\cdot x$ then $S_{y} = g \cdot S_{x}$.  If $y = g' \cdot x$ then $g' =
gh$ for some $h \in H$.  We immediately see that since $g \cdot S_{x}
= g' \cdot S_{x}$ we must have that $h \cdot S_{x} = S_{x}$ for all $h
\in H$.

\subsection{General Case}
\label{sec:group-general}

Assume we have a connected Lie group $G$ acting on $N$ via isometries.
The orbit of $z\in N$ by the $G$ action will be denoted by
$\orbit_{z}$.  If $G_{z}$ is the isotropy group at $z$ then
$\orbit_{z} \approx G/G_{z}$.  We assume all the $G_{z}\subset G$ are
isomorphic as we vary $z\in N$.  The orbits will foliate $N$.  Under
our assumptions, the dimensions of the orbits are constant and we have
a fibration $\pi:N \to M$ such that if $\pi(z)=x$ then the fiber at
$x$ is isomorphic to the orbit $F_{x} \approx \orbit_{z}$.

Let $\dim \orbit_{z} = q$ then the foliation reduces the structure
group of the orthonormal frame bundle from $\SOrth(n)$ to $\SOrth(p)
\times \SOrth(q)$.  Let $y \in \orbit_{z}$ such that $y = g \cdot z$
for $g\in G$.  If we write $T_{z}N = T_{z}\orbit_{z} \oplus T_{z}
\orbit_{z}^{\perp}$ and $T_{y}N = T_{y}\orbit_{z} \oplus T_{y}
\orbit_{z}^{\perp}$ then we have that $g\in G$ takes $T_{z}\orbit_{z}$
isometrically to $T_{y}\orbit_{z}$ and $T_{z}\orbit_{z}^{\perp}$
isometrically to $T_{y}\orbit_{z}^{\perp}$.  The isometric action on
the normal bundle to the orbits tells us that we have a riemannian
submersion and therefore structural equations \eqref{eq:submersion}
are valid.  Additionally we have an isometric action on the fibers.
The Killing vector field is tangential to the orbits therefore its 
lift to the reduced frame bundle is of the form
\begin{equation}
    V = V^{a} \vece_{a} + \half V_{ab} \vece_{ab} + \half 
    V_{ij}\vece_{ij}\,.
    \label{eq:V-Killing}
\end{equation}
Using the invariance conditions $\Lieder_{V} \theta^{i} = 0$, 
$\Lieder_{V}\theta^{a}=0$, and structural equations
\eqref{eq:submersion} leads to
\begin{equation}
    \begin{split}
	V_{ij} & = 0 \,,\\
	V_{a;b} + V_{b;a} &=0\,,\\
	V_{ab} - V_{a;b}& =0 \,,\\
	V_{a;i} - K_{abi}V^{b} &=0\,,
    \end{split}
    \label{eq:Killing-V}
\end{equation}
where $DV_{a} = dV_{a} + \omega_{ab}V_{b} = V_{a;b} \theta^{b} +
V_{a;i}\theta^{i}$.  The first equation tells us that the $G$ action
on $T_{z}\orbit_{z}^{\perp}$ is trivial\footnote{This is one of these
left-right action confusions.  The reader is urged to understand this
in the $S^{2} \approx \SOrth(3)/\SOrth(2)$ example.}.  The middle two
equations tell us that when we restrict to the orbit $\orbit_{z}$ then
we have the familiar Killing equations.  If the orbit bends,
\emph{i.e.}, the second fundamental form is nonvanishing, then the
last equation tell us that the normal derivative of the Killing vector
field is nontrivial and is determined by the second fundamental form.

At $z\in N$ we have that $G_{z}$ acts as an isometry on $T_{z}N
\approx T_{z}\orbit_{z} \oplus T_{z}\orbit_{z}^{\perp}$.  Since
$\orbit_{z}$ is the orbit of $G$, the isotropy subgroup $G_{z}$ acts
as an isometry and leaves invariant the tangent space
$T_{z}\orbit_{z}$.  Consequently, $G_{z}$ also acts as an isometry on
the orthogonal complement $T_{z}\orbit_{z}^{\perp}$.  The action of
$G$ is transitive on the orbit $\orbit_{z}$.  If we look at the
structural equations \eqref{eq:submersion} we see that the second
fundamental tensor $K_{abi}(z)$ and the integrability tensor
$A_{ija}(z)$ must be invariant tensors under the $G_{z}$ action by
generalizing the arguments given in Section~\ref{sec:group-transitive}
to the normal bundle $T_{z}\orbit_{z}^{\perp}$.  Also remember 
$K_{abi}$ and $A_{ija}$ are ordinary functions on the reduced frame 
bundle. The structure equations show that these functions are 
constants under the action of $G$. To see this consider a Killing 
vector $V$ then we note that $\Lieder_{V}(d\theta^{a}) = 
d(\Lieder_{V}\theta^{a}) = 0$ and therefore
\begin{equation*}
    0 = -(\Lieder_{V}\omega_{ab})\wedge\theta^{b} -V(K_{abi}) 
    \theta^{b}\wedge \theta^{i} -V(A_{ija}) 
    \theta^{i}\wedge\theta^{j}\,.
\end{equation*}
Note there is a unique term that is a form of degree $2$ in the
horizontal direction and therefore $V(A_{abi})=0$ and we conclude that
$A_{ija}$ is constant under the action of $G$.  This reduces the
equation to $[\Lieder_{V}\omega_{ab} -
V(K_{abi})\theta^{i}]\wedge\theta^{b} =0$. Cartan's lemma tells us 
that 
\begin{equation*}
    \Lieder_{V}\omega_{ab} - V(K_{abi})\theta^{i} = 
    B_{abc}\theta^{c}\,,
\end{equation*}
where $B_{abc}= B_{acb}$.  Symmetrizing the displayed equation under
$a \leftrightarrow b$ we see that $V(K_{abi})\theta^{i} =
\half(B_{abc}+B_{bac}) \theta^{c}$.  We immediately see that
$V(K_{abi})=0$ and $B_{abc}=-B_{bac}$.  This tells us that $K_{abi}$
is constant under the action of $G$.  Also $B_{abc}$ is skew under $a
\leftrightarrow b$ but symmetric under $b \leftrightarrow c$ and
therefore $B_{abc}=0$ and consequently we also learn as expected
$\Lieder_{V}\omega_{ab}=0$.  The same type of statements will be true
for the curvatures. We summarize below.

\begin{prop} 
    Let $u \in\Redframe(N)$ and let $B_{u} \subset \Redframe(N)$ be
    the orbit\footnote{It can be shown that $B_{u}$ is a sub-bundle of
    $\Redframe(N)$.} of $u$ under the action of $G$.  The functions
    $K_{abi}$, $A_{ija}$ and the curvatures are constant on $B_{u}$.
\end{prop}

We have some information of the derivatives of various tensors.  As an
example we consider the case of the extrinsic curvature. Remember that
$K_{abi}$ are functions on $\Redframe(N)$ and the differential
$dK_{abi}$ in the various directions on $\Redframe(N)$ are given by
\begin{equation}
    dK_{abi}= -\omega_{ac}K_{cbi} -\omega_{bc}K_{aci} -\pi_{ij}K_{abj}
    + K_{abi;j}\theta^{j} + K_{abi;c}\theta^{c}\,.
    \label{eq:dK}
\end{equation}
If we differentiate along the direction of the Killing vector field we
find using \eqref{eq:Killing-V} and the previous equation that
\begin{equation}
     0= -V_{ac}K_{cbi} -V_{bc}K_{aci} + K_{abi;c}V^{c}\,.
    \label{eq:dK-1}
\end{equation}

If the integrability tensor $A_{ija}$ vanishes then the horizontal
distribution is integrable.  Each leaf of the foliation is isometric
to the base $M$ and also each leaf is orthogonal to the fibers.  This
can easily happen because of the group action.  Assume that under the
$G_{z}$ action there are no fixed vectors in $T_{z}\orbit_{z}$.  The
integrability tensor $A_{ij}{}^{a}(z)$ is an invariant tensor under
$G_{z}$ that transforms just like a vector in the vector space
$T_{z}\orbit_{z}$ and thus it must vanish.  This extends everywhere
because we are assuming that the $G$ action leads to a \emph{bona
fide} fibration and the vector spaces $T_{z}\orbit_{z}$, groups
$G_{z}$ and the associated representations are all isomorphic.  This
leads to the following proposition.

\begin{prop}
    If under the $G_{z}$
    action there are no fixed vectors in $T_{z}\orbit_{z}$ then 
    $A_{ija}=0$. The horizontal distribution is integrable and its 
    integral submanifolds are orthogonal to the fibers.
\end{prop}

These methods are also useful for studying axisymmetric solutions.
For example, Theorem~7.1.1 in Wald's book~\cite{Wald:book} can be
proven by using the methods discussed above.

\subsection{The Basic Example}
\label{sec:basic-example}

The basic non-trivial example is given by the $\SOrth(3)$ action on $N
= \bbE^{3}\backslash \{\bm{0}\}$.  The fibration $\pi: N \to
M$ has fibers isomorphic to $S^{2}$ and the base is $M = \bbR_{+}$.
We can easily write down the structural equations
\eqref{eq:submersion} by noting that since $\dim M = 1$ the
integrability tensor $A_{ija}$ vanishes identically.  Note that the
indices $i,j = 1$ and they will be suppressed.  At $z \in N$, the
isotropy group is isomorphic to $\SOrth(2)$ and therefore the second
fundamental form must of the form $K_{ab} = k \delta_{ab}$ where $k$
is constant on each $S^{2}$ fiber and can only depend on the radial
direction, \emph{i.e.}, it is the pullback of a function on $M$.
Let's write the $1$-forms as $(\rho,\theta,\phi)$ where we are using a
notation analogous to spherical coordinates $(r,\vartheta,\varphi)$.
\begin{equation*}
    \begin{split}
	\omega_{ai} &= k \theta^{a}\,,\\
	d\theta &= - \omega \wedge \phi - k \theta \wedge \rho\,,\\
	d\phi &= + \omega \wedge \theta - k \phi \wedge \rho\,,\\
	d\rho & = 0\,.
	\end{split}
\end{equation*}
Next we observe that since $\bbE^{3}$ is flat we have that
$d\omega_{ai} +\omega_{ab} \wedge \omega_{bi} + \omega_{aj} \wedge
\omega_{ji} =0$. This greatly simplifies to $d(k\theta^{a}) + k 
\omega \wedge \theta^{a} =0$. A little algebra yields $(dk + k^{2} 
\rho) \wedge \theta^{a} = 0$ and consequently
\begin{equation}
    dk =- k^{2}\, \rho\,.
    \label{eq:xx-1}
\end{equation}
This immediately tells us that $k$ is constant on the fibers as
expected.  From $d^{2}\theta = d^{2} \phi = 0$ we learn that $d\omega
= A \theta\wedge \phi$.  Note that $\omega$ is invariant on an orbit,
the area element $\theta \wedge \phi$ is invariant on the orbit,
therefore $A$ must be constant on the orbits.  The equation $d^{2}
\omega = 0$ then tells us that $dA + 2kA\rho = 0$.

Since $d\rho = 0$ we can set $\rho = dr$ for some function $r$ 
specified up to an additive constant. We can easily integrate 
\eqref{eq:xx-1} to obtain $1/k  = r + c$. By redefining the 
coordinate $r$ we can set $c=0$. Thus we find that the second 
fundamental form is determined by
\begin{equation}
    k = \frac{1}{r}.
    \label{eq:k-soln}
\end{equation}
The curvature is given by
\begin{equation}
    d\omega = \frac{A_{0}}{r^{2}}\; \theta \wedge \phi\,,
    \label{eq:A-soln}
\end{equation}
where $A_{0} \in \bbR$.  $A_{0}$ may be determined by topological or
by geometrical considerations.

In summary the structural equations associated with the $\SOrth(3)$ 
action on $\bbE^{3} \backslash \{\bm{0}\}$ become
\begin{equation}
    \begin{split}
	\rho & = dr\,, \\
	d\theta &= - \omega \wedge \phi - \frac{1}{r}\, \theta \wedge dr\,,\\
	d\phi &= + \omega \wedge \theta - \frac{1}{r}\, \phi \wedge dr\,,\\
	d\omega & = \frac{A_{0}}{r^{2}}\; \theta \wedge \phi\,.
    \end{split}
    \label{eq:st-r3}
\end{equation}
On a level surface of constant $r$ we have that $(\theta,\phi,\omega)$
are the Maurer-Cartan forms for $\SOrth(3)$.  The constant $A_{0}$ may
be determined via topological arguments or coordinate arguments where
you find that $A_{0} = 1$.

The space $N=\bbE^{3}\backslash \{\bm{0}\}$ is flat and the geodesics
are straight lines.  It is not a complete riemannian manifold because
radial geodesics reach the origin in finite affine parameter.  It is
clear that by adding an extra point $N \cup \{\bm{0}\} \approx
\bbE^{3}$ becomes a complete riemannian manifold.  We can try to do
the same analysis by thinking of $N$ in terms of its structural
equation \eqref{eq:st-r3}.  This analysis is much more complicated.
The structural equations are non-singular as long as $r\neq 0$ and
this is true in $N$.  Analyzing the structural equation you can see
that radial geodesics get to $r=0$ in finite time and thus it appears
that there is a hole in the space.  You have to work little hard with
the structural equations to show that the apparent singularity at
$r=0$ is removable and that by adding a point at $r=0$ we get the
complete smooth manifold $\bbE^{3}$.

\section{Spherically Symmetric $(3+1)$ Geometry}
\label{sec:birkhoff}

Assume $N$ is a $4$-dimensional lorentzian manifold that is both
orientable and time orientable.  This means that the structure group
of of the orthonormal Lorentz frame bundle is
$\SOrth^{\uparrow}(1,3)$, the connected component of the Lorentz
group.  We assume there is an $\SOrth(3)$ action that leaves the
metric invariant and that the orbit of a point is a $2$-dimensional
spacelike surface.  Let $\orbit_{p}$ be the orbit through $p \in N$.
This action leads to a foliation of $N$ by the $2$-dimensional orbits.
Under some assumptions of a constant dimensionality of the orbits we
can assume that this foliation is actually a fibration.  Our
hypothesis tells us that $\dim\orbit_{p} = 2$.  If $G_{p}$ is the
isotropy group at $p$ then $\dim G_{p} = 1$.  This tells us that
$\orbit_{p} \approx \SOrth(3)/G_{p} \approx S^{2}$.  The $\SOrth(3)$
action identifies a spacelike $2$-plane $T_{p}\orbit_{p} \subset
T_{p}N$ at each $p\in N$ that induces a reduction of the structure
group of the lorentzian orthonormal frame bundle from
$\SOrth^{\uparrow}(1,3)$ to $\SOrth^{\uparrow}(1,1) \times \SOrth(2)$.
A consequence is that there is only one $\omega_{ab}$ and one
$\pi_{ij}$.  If we use some type of ``Schwarzschild spherical
coordinates'' denoted by $(t,r,\vartheta,\phi)$.  Then we will have
non-vanishing connections $\omega_{\vartheta\phi}$ and $\pi_{tr}$.  If
$\pi:N\to M$ is our fiber bundle and if $\pi(p) = x$ then the fiber
over $x$ is given by $F_{x} = \orbit_{p}$.

The general discussion of Section~\ref{sec:group-general} tells us
that we have a pseudo-riemannian submersion.  At $p \in N$ we can
write $T_{p}N = T_{p}\orbit_{p} \oplus T_{p}\orbit_{p}^{\perp}$ and
the $\SOrth(3)$ action tells us that both the riemannian metric on
$T_{p}\orbit_{p}$ and the lorentzian metric on
$T_{p}\orbit_{p}^{\perp}$ are invariant under the $\SOrth(3)$ action.
At $p \in N$, all geometrical structures must be invariant under the
isotropy group action $G_{p} \approx \SOrth(2)$.  The action of
$G_{p}$ on $T_{p} \orbit_{p}$ is the standard $\SOrth(2)$ action and
the action on $T_{p}\orbit_{p}^{\perp}$ is automatically trivial
because there is no $\SOrth(2)$ subgroup in $\SOrth^{\uparrow}(1,1)$.
Because of this we can conclude that the integrability tensor
$A_{ija}$ for the horizontal spaces of the submersion must vanish.  At
$p \in N$ the integrability tensor may be viewed as a map $A_{p}:
\Lambda^{2}(T_{p}\orbit_{p}^{\perp}) \to T_{p}\orbit_{p}$.  We note
that $\Lambda^{2}(T_{p}\orbit_{p}^{\perp})$ is one dimensional so
$A_{p}$ on the normalized area element gives a preferred vector on
$T_{p}\orbit_{p}$.  Said differently we must have $A_{ija} =
\epsilon_{ij}v_{a}$.  A non-vanishing vector field $v$ tangential to
$\orbit_{p}$ is not $G_{p}$ invariant and must therefore
vanish\footnote{The tensor $A_{ija}$ must be invariant under $G_{p}$
if not then the structure group will be further reduced.}.  This tells
us that $v$ must vanish.  Thus we have learned that the horizontal
subspaces are integrable.

Similarly the second fundamental forms $K_{abi}(p)$ must be invariant
under $G_{p}$ otherwise the structure group will be further reduced.
This immediately tells us that $K_{abi} = \delta_{ab} S_{i}$ where
$\sigma = S_{i}\theta^{i}$ will be called the \emph{second fundamental
$1$-form}.

We can extend this argument and conclude that $R^{N}_{ia} =0$ and
$R^{N}_{ab} \propto \delta_{ab}$.  We will denote by  $\eta = \bigl(
\begin{smallmatrix} -1 & 0 \\ 0 & +1 \end{smallmatrix} \bigr)$  the
Minkowski metric on $T\orbit_{p}^{\perp}$.

The structural equations \eqref{eq:submersion} applied to this case
become
\begin{equation}
    \begin{split}
	\omega_{ai} &= S_{i}\theta^{a}\,,  \\
	\pi_{ij} &= \omega_{ij}\,, \\
	d\theta^{i} &= -\pi_{ij}\wedge\theta^{j}\,,  \\
	d\theta^{a} &= -\omega_{ab}\wedge\theta^{b} 
	-S_{i}\theta^{a}\wedge\theta^{i}\,.
    \end{split}
    \label{eq:spherical}
\end{equation}

Next we explore additional properties that follow from the $\SOrth(3)$
action.  First we observe that $\omega_{ab}$ did not get modified by
the symmetry breakdown therefore we know that under the action of the
$\SOrth(3)$ Killing vector
\begin{equation}
    V = V^{a} \vece_{a} + \half V_{a;b} \vece_{ab}\,,
    \label{eq:Killing-V-exp}
\end{equation}
the connection is invariant $\Lieder_{V}\omega_{ab} = 0$.  Also
$\Lieder_{V} \omega_{ai} = 0$ because $\omega_{ai}$ in
\eqref{eq:spherical} comes from restriction to the reduced orthonormal
frame bundle $\Redframe(N)$ and the Killing vector field $V$ is
tangential to $\Redframe(N) \subset \Orthframe(N)$.  This immediately
implies that $\interior_{V} dS_{i} = S_{i;a}V^{a}=0$ from which we
conclude that $S_{i;a}=0$.  Next we show that the $1$-form $\sigma =
S_{i} \theta^{i}$ is invariant under the $\SOrth(3)$ action.  We have
that $\Lieder_{V} \sigma = \interior_{V}d\sigma + d
\interior_{V}\sigma$.  We observe that $d\sigma = S_{i;j}\theta^{j}
+ S_{i;a}\theta^{a} = S_{i;j}\theta^{j}$.  Since $\interior_{V} \sigma
= 0$ and $\interior_{V} d\sigma = 0$ we see that
$\Lieder_{V}\sigma=0$.

Alternatively, this can also be seen by looking at the reduced
structure equations \eqref{eq:spherical} directly.  First we observe
that $\Lieder_{V} \sigma = \Lieder_{V}(S_{i}\theta^{i}) =
(\interior_{V}dS_{i}) \theta^{i} = V^{a}S_{i;a}\theta^{i}$.  Next we
note that $0 = d (\Lieder_{V}\theta^{a})= \Lieder_{V}(d\theta^{a})$
from which we conclude that $0 = (\Lieder_{V}\omega_{ab}) \wedge
\theta^{b} + (\Lieder_{V}\sigma) \wedge \theta^{a}$.  If we write
$\omega_{ab} = \epsilon_{ab} \omega$ we see that the previous equation
becomes
\begin{equation*}
    \begin{split}
       (\Lieder_{V}\omega) \wedge \theta^{3} + V^{a}S_{i;a}\theta^{i} 
       \wedge \theta^{2}  &=0\,, \\
       -(\Lieder_{V}\omega) \wedge \theta^{2} + V^{a}S_{i;a}\theta^{i} 
       \wedge \theta^{3}  &=0\,.
    \end{split}
\end{equation*}
By inspection we see that the unique solution is $\Lieder_{V} \omega = 
0$ and $S_{i;a}=0$.
In conclusion we have that
\begin{equation}
    \begin{split}
	d\sigma &= - \half(S_{i;j} - S_{j;i}) \theta^{i} \wedge
	\theta^{j}\,,\\
	S_{i;a} &= 0\,.
    \end{split}
    \label{eq:extd-s}
\end{equation}
Notice that \eqref{eq:A-K-der} tells us that $S_{i;j}=S_{j;i}$ and 
therefore we learn that $d\sigma=0$.
This is also a consequence of
\begin{equation*}
    0 = d^{2}\theta^{a} = - \left( \epsilon_{ab} \omega + 
    \delta_{ab}d\sigma \right) \wedge \theta^{b}\,.
\end{equation*}
One of the equations above is $\omega \wedge \theta^{3} + d\sigma
\wedge\theta^{2} = 0$.  If we wedge with $\theta^{3}$ we find that
$d\sigma\wedge \theta^{2}\wedge \theta^{3}=0$.  If we use
\eqref{eq:extd-s} we immediately learn that $S_{0;1}=S_{1;0}$,
\emph{i.e.}, $\sigma$ is a closed $1$-form: $d\sigma=0$. Therefore we 
see that $d\omega \wedge \theta^{a}=0$. Using a similar argument we 
see that
\begin{equation}
    d\omega = k^{F_{x}}\theta^{2} \wedge \theta^{3}\,.
    \label{eq:r-ab-1}
\end{equation}
We will shortly return to this equation.

Using the submersion curvature results in
Table~\ref{tbl:curv-submersion} we immediately see that
\begin{align}
    R^{N}_{ijkl} & = R^{M}_{ijkl}\,,
    \label{eq:r-ijkl}  \\
    R^{N}_{ijab} & = 0\,,
    \label{eq:r-ijab}  \\
    R^{N}_{ijka} & = 0\,.
    \label{eq:r-ijka} \\
    R^{N}_{aibj} &= - \delta_{ab}(S_{i;j} + S_{i}S_{j})\,.
    \label{eq:r-aibj}
\end{align}
Note that the $G_{p}$ action implies that $R^{N}_{abci} =
R^{N}_{ciab}=0$ and this can explicitly be verified from the formulas.
We point out that
\begin{equation}
    R^{M}_{ijkl} = k^{M} \epsilon_{ij} \epsilon_{kl} =
    -k^{M}(\eta_{ik}\eta_{jl} - \eta_{il}\eta_{jk}) \,,
    \label{eq:RM}
\end{equation}
because $M$ is two dimensional.  The negative sign in the last
equality is due to the negative sign in the Minkowski metric.

We observe that since the fiber $\orbit_{p}$ is $2$-dimensional we
have that $R^{F_{x}}_{abcd} =k^{F_{x}} \epsilon_{ab}\epsilon_{cd} =
k^{F_{x}}(\delta_{ac}\delta_{bd}- \delta_{ad}\delta_{bc})$.  Again using 
the results from Table~\ref{tbl:curv-submersion} we see that
\begin{equation}
    R^{N}{}_{abcd} = (k^{F_{x}} - S^{i}S_{i})(\delta_{ac}\delta_{bd}  
    -\delta_{ad}\delta_{bc})\,.
    \label{eq:gauss-sch}
\end{equation}
Putting all this together we learn that the Cartan structural equation
for the $\SOrth(2)$ curvature \eqref{eq:cartan-vert-1} may be written
in this case as
\begin{equation}
    d\omega_{ab} = k^{F_{x}}\theta^{a} \wedge \theta^{b}\,.
    \label{eq:r-ab}
\end{equation}
The $\SOrth(3)$-orbits are ``round'' $2$-spheres and we have that
\begin{equation}
    k^{F_{x}} = \frac{1}{r^{2}}\,.
    \label{eq:kF-r2}
\end{equation}
Here $r:\Redframe(N)\to\bbR_{+}$ is the radius of the $2$-sphere.
Since know that $k^{F_{x}}$ is constant on each orbit there exists a
globally defined function $r_{M}:M\to \bbR_{+}$ such that $r$ is the
pullback to the bundle of $r_{M}$.  The function $r_{M}$ is just the
radius of the fibering $S^{2}$.  We will avoid all the notation
required and simply refer to the radius function as $r$ and implicitly
assume its domain on context.  Note that $dk^{F_{x}}$ is independent
of $\theta^{a}$ because of the $\SOrth(3)$ action and it is also
independent of the connections because it is invariant under
$\SOrth^{\uparrow}(1,1) \times \SOrth(2)$.  Computing $0 = d^{2}
\omega_{ab}$ we find
\begin{equation*}
    0 = \left(dk^{F_{x}}+ 2 k^{F_{x}}\, \sigma \right)
    \wedge \theta^{2} \wedge \theta^{3}\,.
\end{equation*}
Since $k^{F_{x}}$ and $\sigma$ are $\SOrth(3)$ invariant we learn that
\begin{equation}
    dk^{F_{x}} + 2 k^{F_{x}}\, \sigma =0\,.
    \label{eq:dkF}
\end{equation}
Using \eqref{eq:kF-r2} we see that
\begin{equation}
    \sigma = \frac{1}{r}\;dr = d(\log r)\,.
    \label{eq:sigma-dr2}
\end{equation}
We have learned that $\sigma$ is exact as will be collaborated by an 
independent argument later. In fact, the above equation will be valid 
everywhere if the fibration is non-singular. Next we observe that if 
$dr = r_{i} \theta^{i}$ then
\begin{equation}
    S_{i} = \frac{r_{i}}{r}\,.
    \label{eq:r-S-rel}
\end{equation}
The Gauss equation \eqref{eq:gauss-sch} becomes
\begin{equation}
	R^{N}{}_{abcd} = \left(\frac{1 - \lVert dr
	\rVert_{M}^{2}}{r^{2}}\right) (\delta_{ac}\delta_{bd}
	-\delta_{ad}\delta_{bc})\,.
    \label{eq:Schw-Gauss}
\end{equation}

The Ricci tensor is computed using \eqref{eq:ricci-submersion}.  First
we note that as expected by $G_{p}$ invariance we have $R^{N}_{ai}=0$.
Doing the computations we find
\begin{equation}
    \begin{split}
        R^{N}_{ij} &= -2(S_{i;j} + S_{i}S_{j}) - k^{M} \eta_{ij}\,,\\
	& = -2 \frac{r_{i;j}}{r} - k^{M} \eta_{ij}\,,\\
        R^{N}_{ab} &=  \left[k^{F} -  (S^{i}{}_{;i} + 2S^{i}S_{i}) \right] 
        \delta_{ab}\, \\
	& = \left( \frac{1 - \lVert dr \rVert^{2}_{M} - r r^{i}{}_{;i}}{r^{2}}
	\right) \delta_{ab}\,,\\
	&= \left( \frac{2 - \Box(r^{2})}{2r^{2}}\right) \delta_{ab}\,.
    \end{split}
    \label{eq:Sch-Ricc}
\end{equation}
The wave operator is defined by $\Box f = \eta^{ij} f_{;i;j}$ where 
each semi-colon denotes a covariant derivative.

The Cartan structural equations associated with the $\SOrth(3)$ action 
on $N$ are
\begin{align}
    d\theta^{0} & = +\pi \wedge \theta^{1}\,,
    \label{eq:SO3-dtheta0}  \\
    d\theta^{1} & = +\pi \wedge \theta^{0}\,,
    \label{eq:SO3-dtheta1} \\
    d\pi & = k^{M}\; \theta^{0} \wedge \theta^{1}\,,
    \quad\text{where}\quad \pi= \pi_{01}\,,
    \label{eq:SO3-dpi} \\
    d\theta^{2} & = - \omega \wedge \theta^{3} - \frac{1}{r}\;
    \theta^{2} \wedge dr\,, 
    \label{eq:SO3-dtheta2} \\
    d\theta^{3} & = + \omega \wedge \theta^{2} - \frac{1}{r}\;
    \theta^{3} \wedge dr\,, 
    \label{eq:SO3-dtheta3} \\
    d\omega & = +\frac{1}{r^{2}}\; \theta^{2} \wedge \theta^{3}\,,
    \quad\text{where}\quad \omega= \omega_{23}\,.
    \label{eq:SO3-domega}
\end{align}
The equation $d^{2}\pi=0$ tells you that $dk^{M} = k^{M}_{1}
\theta^{1} + k^{M}_{2} \theta^{2}$, \emph{i.e.}, $k^{M}$ is the
pullback to the frame bundle of a function on $M$.  The geometry is
determined by two functions, $r$ and $k^{M}$, that are the pullbacks
of functions on $M$.  If $r$ and $k^{M}$ are non-singular in a
neighborhood of a point $q\in \Redframe(N)$ then the structural
equations can be integrated to locally construct the frame bundle.

If $r$ and $k^{M}$ are independent functions in a
neighborhood in $M$, \emph{i.e.}, $dr \wedge dk^{M} \neq 0$, then the
inverse function theorem tells you that $\varphi: p \in M \mapsto
(r(p),k^{M}(p)) \in \bbR^{2}$ can be used as a local coordinate system
for the neighborhood.

The converse of the above will be important to us later. 
If $r$ and $k^{M}$ are dependent functions in a
neighborhood in $M$, \emph{i.e.}, $dr \wedge dk^{M} = 0$, then 
$k^{M}$ is a function of $r$. The reason is that $d(k^{M}\,dr)=0$ 
and therefore locally there exists a function $F$  such that $dF = 
k^{M}\,dr$.

\section{Vacuum Einstein Equations}
\label{sec:vac-E-eqs}

The vacuum Einstein equations are $R^{N}_{ij}=0$ and 
$R^{N}_{ab} = 0$. Using \eqref{eq:Sch-Ricc} these may be written as
\begin{equation}
    \begin{split}
        -2\frac{r_{i;j}}{r} - k^{M} \eta_{ij}
	 & = 0 \,,\\
	 \left( \frac{1 - \lVert dr \rVert^{2}_{M} - r
	 r^{i}{}_{;i}}{r^{2}} \right) \delta_{ab} &=0\,.
    \end{split}
    \label{eq:Einstein-eqs}
\end{equation}
Taking the trace of each of the equations above we learn
\begin{equation}
    \begin{split}
	\frac{r^{i}{}_{;i}}{r} + k^{M} &= 0\,,\\
	\frac{r^{i}{}_{;i}}{r} - \frac{1 - \lVert dr 
	\rVert^{2}_{M}}{r^{2}} &= 0\,.
    \end{split}
    \label{eq:E-trace}
\end{equation}
Taking the difference of the equations above we see that
\begin{equation}
    k^{M} =  -\frac{1}{r^{2}} \left( 1 - \lVert dr
    \rVert^{2}_{M} \right)\,.
    \label{eq:curv-MN-rel}
\end{equation}
The dalembertian term of \eqref{eq:E-trace} may be rewritten as
\begin{equation}
    \Box (r^{2}) =2\;,
    \label{eq:box-r}
\end{equation}
a hyperbolic equation for $r^{2}$.  In some sense, the area of the
fibering $2$-sphere, $A=4\pi r^{2}$, is a propagating field on $M$
with a constant source.

Finally we point out that automatically there is an extra killing 
vector. Consider a ``horizontal'' vector field
\begin{equation*}
    X = X^{i} \vece_{i} + \half X^{ij} \vece_{ij}
\end{equation*}
then we have
\begin{equation}
    \begin{split}
	\Lieder_{X}\theta^{a} &= 
	\frac{X^{i}r_{i}}{r} \theta^{a}\,, \\
        \Lieder_{X} \theta^{i} &= -X_{ij}\theta^{j} + 
	DX^{i}\,.
    \end{split}
    \label{eq:Lie-met}
\end{equation}
Notice that for any horizontal $X$, the change in $\theta^{a} \otimes
\theta^{a}$ is conformal\footnote{This has to be true because there is
a unique round metric on $S^{2}$ up to scale.  The horizontal vector
field moves you to another point where the associated fiber is also a
round $S^{2}$.}.  On the other hand if $X^{i} \propto
\epsilon^{ij}r_{j}$ then the first equation above is automatically
zero.  We will see that we can make the second also zero.  Choose
$X^{i} = \epsilon^{ik}r_{k} F(r)$ for some real valued function of
$r$.  We note that $DX_{i} = X_{i;j}\theta^{j} + X_{i;a} \theta^{a}$.
The Killing conditions require $X_{i;a}=0$, \emph{i.e.}, $X$ is
intrinsically associated with the base $M$.  The second displayed
equation above also requires $X_{i;j} + X_{j;i} = 0$.  Next we note
that
\begin{equation*}
    \begin{split}
    X_{i;j}  &= \epsilon_{i}{}^{k}r_{k;j}F(r) + \epsilon_{i}{}^{k}
    r_{k} r_{j}\, F'(r)\,, \\
    & = \epsilon_{ik} r^{k}r_{j} F'(r) - \half \epsilon_{ij}k^{M} r
    F(r)\,.
    \end{split}
\end{equation*}
The condition for the flow to generate an isometry is $F'(r)=0$ or
equivalently $F(r) = F_{0}$ where $F_{0}$ is a constant.  In
conclusion we have an additional Killing vector given by
\begin{equation}
    X^{i} = - \epsilon^{ij} r_{j}\,.
    \label{eq:Killing-extra}
\end{equation}
The vectors $r^{i}$ and $X^{i}$ are Minkowski orthogonal, $r_{i} X^{i}
=0$, and that
\begin{equation}
    \left\lVert X \right\rVert^{2}_{M} = - \left\lVert \bnabla r
    \right\rVert^{2}_{M}\,.
    \label{eq:S-X-norm}
\end{equation}
Note that if $\bnabla r$ is
spacelike then $X$ is timelike and vice-versa.  If $\bnabla r$ is lightlike
then $X$ is also lightlike and vice-versa. 

Next we observe that the Lie derivative of the metric on the fibers
along the direction of $\bnabla r$ is given by
\begin{equation}
    \Lieder_{\bnabla r}(\theta^{a} \otimes \theta^{a}) = 2
    \frac{\left\lVert \bnabla r \right\rVert^{2}_{M}}{r}\,
    \theta^{a} \otimes \theta^{a}\,.
    \label{eq:Lie-along-S}
\end{equation}
We conclude that if $\bnabla r$ is space-like then the area of the
$2$-sphere increases in the direction of $\bnabla r$.  If $\bnabla r$
is time-like then the area of the $2$-sphere decreases in the
direction of $\bnabla r$.

\subsection{Properties of the radius function}
\label{sec:radius-function}

Next we derive a differential equation satisfied by $\nu=r^{i}r_{i} =
\lVert dr \rVert^{2}_{M}$.
\begin{equation*}
    \begin{split}
	d \nu &= 2 r^{i}r_{i;j}\theta^{j}\,, \\
	 &= -r k^{M}\,dr\,, \\
	 &= \frac{1-\nu}{r}\,dr\,.
    \end{split}
\end{equation*}
A little algebra leads to the equation
\begin{equation}
    d\bigl((\nu-1)r\bigr) =0\,.
    \label{eq:r-S}
\end{equation}
The solution to this equation is elementary and given by
\begin{equation}
    \nu = \lVert dr \rVert^{2}_{M} = 1 + \frac{c}{r} \,,
    \label{eq:S2-sol}
\end{equation}
where $c \in \bbR$ is a constant of integration.
We also have using
\eqref{eq:S-X-norm}
\begin{equation}
    \left\lVert X \right\rVert^{2}_{M} = -\left\lVert dr
    \right\rVert^{2}_{M} = 
    - \left( 1 + \frac{c}{r} \right) .
    \label{eq:S-X-norm-1}
\end{equation}
Using \eqref{eq:curv-MN-rel} we see that
\begin{equation}
     k^{M} =  \frac{c}{r^{3}}\,.
    \label{eq:curm-MN-rel-1}
\end{equation}

Next we determine the constant $c$.  Here we need to make a physical
assumption.  We assume that in the spacetime $N$ there is a region
that is asymptotically minkowskian and looks like the gravitational
far field of a localized mass distribution.  The Cartan structural
equations tell us that as $r \to +\infty$ our geometry becomes
asymptotically minkowskian.  The equation for geodesic deviation says
that $D_{u}D_{u}\eta = D_{u}D_{\eta}u = [D_{u},D_{\eta}]u =
R^{N}(u,\eta)u$.  In the instantaneous rest frame we have $u =
\vece_{t}$ and we look at $\eta = \eta^{r}\vece_{r}$.  Our relative
radial acceleration equation becomes
\begin{equation*}
    \frac{d^{2} \eta^{r}}{dt^{2}} = R^{r}{}_{ttr}\eta^{r} = 
    R_{rttr}\eta^{r} = -k^{M}\eta^{r}  =- \frac{c}{r^{3}}\;\eta^{r}\,.
\end{equation*}
Newtonian mechanics tells us that $\ddot{r}  = -M/r^{2}$ where $M$ 
is the mass of the star. We have that $\eta^{r} = \delta r$ and 
therefore $\ddot{\eta}^{r} = (2M/r^{3}) \eta^{r}$. We immediately 
see that
\begin{equation}
    c = -\RS \qquad\text{where}\qquad \RS=2M
    \label{eq:RS}
\end{equation}
is called the Schwarzschild radius.

Finally we observe that we can define a closed $1$-form $\tau$
by\footnote{On a two dimensional manifold, a locally non-vanishing
$1$-form $\alpha$ always defines a local foliation because the
Frobenius condition $d\alpha \equiv 0 \mod \alpha$ is automatically
satisfied.  Furthermore, the Frobenius theorem states that there
exists functions $f$ and $g$ such that $\alpha = f\,dg$.  In our case
we have that $\alpha = *dr$.}
\begin{equation}
    \tau = \frac{r_{1}\theta^{0} + r_{0}\theta^{1}}{1-\RS/r}
    = \frac{-\epsilon_{ij}r^{i}\theta^{j}}{1-\RS/r}\,,
    \label{eq:def-tau}
\end{equation}
with the property that $\tau(X)=1$. Note that $\tau$ is not defined 
if $r=\RS$. Up to scale we have that $\tau$ is basically $*dr$, the 
Hodge dual on M of $dr$.

It is worthwhile to summarize the data that determines our
geometry:
\begin{align}
    \RS & = 2M\,, \quad\text{Schwarzschild radius},
    \label{eq:BH-RS}  \\
    r_{M} & : M \to \bbR_{+}\,, \quad\text{radius of the fibering } S^{2}\,,
    \label{eq:BH-r}  \\
    dr & = r_{i} \theta^{i}\,,
    \label{eq:BH-dr} \\
    r_{i;j}  &= \frac{\RS}{2 r^{2}}\; \eta_{ij}\,,
    \label{eq:BH-rij}\\
    X^{i} & =  - \epsilon^{ij}r_{j}\,,
    \label{eq:BH-X}  \\
    \lVert dr \rVert^{2}_{M} & = 1 - \frac{\RS}{r} \,,
    \label{eq:BH-dr2} \\
    \lVert X \rVert^{2}_{M} & = -\left( 1 - 
    \frac{\RS}{r} \right)\,,
    \label{eq:BH-X2}  \\
    \tau & = \frac{r_{1}\theta^{0} + r_{0}\theta^{1}}{1-\RS/r} =
    \frac{-\epsilon_{ij}r^{i}\theta^{j}}{1-\RS/r}\;\text{ and } \;
    d\tau=0\,, \label{eq:BH-tau} \\
    \lVert \tau \rVert^{2}_{M} & = -\left(1 -
    \frac{\RS}{r}\right)^{-1} \,.
    \label{eq:BH-tau2}
\end{align}

The Cartan structural equations for the reduced frame bundle of the
Schwarzschild spacetime are
\begin{align}
    d\theta^{0} & = +\pi \wedge \theta^{1}\,,
    \label{eq:BH-dtheta0}  \\
    d\theta^{1} & = +\pi \wedge \theta^{0}\,,
    \label{eq:BH-dtheta1} \\
    d\pi & = - \frac{\RS}{r^{3}}\; \theta^{0} \wedge \theta^{1}\,,
    \quad\text{where}\quad \pi= \pi_{01}\,,
    \label{eq:BH-dpi} \\
    d\theta^{2} & = - \omega \wedge \theta^{3} - \frac{1}{r}\;
    \theta^{2} \wedge dr\,, 
    \label{eq:BH-dtheta2} \\
    d\theta^{3} & = + \omega \wedge \theta^{2} - \frac{1}{r}\;
    \theta^{3} \wedge dr\,, 
    \label{eq:BH-dtheta3} \\
    d\omega & = +\frac{1}{r^{2}}\; \theta^{2} \wedge \theta^{3}\,,
    \quad\text{where}\quad \omega= \omega_{23}\,.
    \label{eq:BH-domega}
\end{align}
Note that the only singularity in the structural equations occurs
where $r=0$.  For this reason we expect the frame bundle of the
Schwarzschild manifold to be smooth everywhere as long as $r\neq 0$.
In particular we do not expect any type of singularity when
$r_{M}=\RS$.  The exceptional properties of the Schwarzschild solution
at $r_{M}=\RS$ occur because of the behavior of $dr$ at $r=\RS$.  The
only potential problems with the structural equations occur at $r=0$.
What type of singularity is at $r=0$?  Is it removable as in the
example of $\bbE^{3} \backslash \{\bm{0}\}$ or is it a true
singularity?

We can make a consistency check on equations\footnote{This also
applies to \eqref{eq:st-r3}.} \eqref{eq:BH-dtheta2},
\eqref{eq:BH-dtheta3}, \eqref{eq:BH-domega}.  If we make a conformal
rescaling $\hat{\theta}^{a} = \theta^{a}/r$ then these equations may
be written as
\begin{align*}
    d\hat{\theta}^{2} & = - \omega\wedge \hat{\theta}^{3}\,,  \\
    d\hat{\theta}^{3} & = + \omega\wedge \hat{\theta}^{2}\,,  \\
    d\omega & = + \hat{\theta}^{2} \wedge \hat{\theta}^{3}\,.
\end{align*}
These equations are easily identifiable.  They are the Cartan
structural equations for the orthogonal frame bundle of the unit
$2$-sphere.  Note that they are the Maurer-Cartan equations for the
group $\SOrth(3)$ and thus the frame bundle of $S^{2}$ is isomorphic
to $\SOrth(3)$. The base space for this frame bundle is precisely 
$\SOrth(3)/\SOrth(2) \approx S^{2}$.

\subsection{Geodesics}
\label{sec:geodesics-1}

We work out some properties of the geodesics on $M$ by using Cartan's
method~\cite{Cartan:riemann}, see Appendix~\ref{sec:geodesics-cartan}.
It is useful to introduce a null basis for the canonical $1$-forms on
the Lorentz frame bundle of $M$ by defining $\theta^{\pm} = \theta^{0}
\pm \theta^{1}$ then the pullback of the metric on $M$ to the frame
bundle is given by $-\half\left( \theta^{+} \otimes \theta^{-} +
\theta^{-} \otimes \theta^{+} \right)$.  We also note that
\begin{equation}
    \begin{split}
    d\theta^{\pm} &= \pm\pi \wedge \theta^{\pm}\,,\\
    d\pi&=  \frac{\RS}{2r^{3}}\; \theta^{+}\wedge \theta^{-}\,.
    \end{split}
    \label{eq:C-lc}
\end{equation} 
Remember that we are working ``upstairs''!

Let $\theta^{i} = \bar{\theta}^{i} + u^{i}d\lambda$ and $\pi =
\bar{\pi}$ for $\lambda \ge 0$.  Here barred $1$-forms are independent
of $d\lambda$ analogous to $\vartheta$ and $\varpi$ in
Appendix~\ref{sec:geodesics-cartan}.  The initial conditions are that
$\bar{\theta}^{i}(0)=0$, $\partial_{\lambda}\bar{\theta}^{i}(0) =
du^{i}$, and $\bar{\pi}(0)=0$.  Differentiating once we see that
\begin{equation*}
    \begin{split}
    \frac{\partial\bar{\theta}^{\pm}}{\partial \lambda} &=
    du^{\pm} \mp \bar{\pi} u^{\pm},,\\
    \frac{\partial \bar{\pi}}{\partial \lambda} & =
    \frac{\RS}{2r^{3}} \left( u^{+}\bar{\theta}^{-} - 
    u^{-}\bar{\theta}^{+} \right).
    \end{split}
\end{equation*}
Differentiating again we see that
\begin{equation}
    \begin{pmatrix}
	\partial^{2} \bar{\theta}^{+}/\partial \lambda^{2}  \\
	\partial^{2} \bar{\theta}^{-}/\partial \lambda^{2}
    \end{pmatrix}
    = \frac{\RS}{2\,r^{3}}
    \begin{pmatrix}
	u^{+}u^{-} & -(u^{+})^{2}  \\
	-(u^{-})^{2} & u^{+}u^{-}
    \end{pmatrix}
    \begin{pmatrix}
	\theta^{+} \\
	\theta^{-}
    \end{pmatrix}   
    \label{eq:ODE-thpm}
\end{equation}

Next we derive an ODE that $r$ satisfies along a geodesic. We note 
that $dr/d\lambda = r_{+}u^{+} + r_{-}u^{-}$. Next we remember that $dr_{\pm} = 
\mp \pi r_{\pm} +r_{\pm;+}\theta^{+} + r_{\pm;-}\theta^{-}$ and that 
$r_{+;+}=r_{-;-}=0$ by \eqref{eq:BH-rij}. Therefore along a geodesic 
we have that
\begin{equation}
    \begin{split}
	\frac{d r_{+}}{d \lambda} &=  
	-\frac{\RS}{4r^{2}}\, u^{-}\,, \\
	\frac{d r_{-}}{d \lambda} &=
	-\frac{\RS}{4r^{2}}\, u^{+}\,.
    \end{split}
    \label{eq:dr-evolution}
\end{equation}
We immediately see that
\begin{equation}
    \frac{d^{2}r}{d\lambda^{2}} = \frac{\RS}{2r^{2}}\, \left\lVert u 
    \right\rVert_{M}^{2}.
    \label{eq:r-ODE}
\end{equation}
The equations that describe the exponential map \eqref{eq:ODE-thpm} 
are complicated but the equation that describes the evolution of $r$ 
along a geodesic \eqref{eq:r-ODE} is relatively simple.

The case of a null radial geodesic is particularly simple because
$d^{2}r/d\lambda^{2}=0$.  If the horizontal lift of the null geodesic
begins at a point $p\in \Orthframe(M)$ with $r(p) = r_{p}$ and $dr(p)
= r_{i}(p) \theta^{i}(p)$ then the evolution of $r$ along the lift is
\begin{equation}
    r(\lambda) = r_{p} + \lambda\left( r_{+}(p) u^{+} + r_{-}(p) u^{-} 
    \right) .
    \label{eq:soln-ODE-r}
\end{equation}
There are four cases of null geodesics to analyze corresponding to
\begin{equation*}
    (u^{+},u^{-}) \in \left\{ (+1,0),(0,+1),(-1,0),(0,-1) \right\}\,.
\end{equation*}
The latter two cases may be considered with the first two by allowing
$\lambda$ to be negative.  In the first case we have that $r(\lambda)
= r_{p} + \lambda r_{+}(p)$, and in the second case we have
$r(\lambda) = r_{p} + \lambda r_{-}(p)$.  Choose a Lorentz frame
$p\in\Orthframe(M)$, if $r_{+}(p)>0$ then $r_{+}(p') >0$ for all $p'$
in the same fiber because the action of the $(1+1)$ dimensional
Lorentz group translates to an action $r_{\pm} \to e^{\pm\eta}r_{\pm}$
where $\eta$ is the rapidity.  This means that we can define the
following four open subsets of $M$:
\begin{equation}
    \begin{split}
        U_{\text{I}} &= \{ q \in M \;|\; r_{+}(p)>0, r_{-}(p) <0\}\,, \\
	U_{\text{II}} &= \{ q \in M \;|\; r_{+}(p)<0, r_{-}(p) <0\}\,, \\
	U_{\text{III}} &= \{ q \in M \;|\; r_{+}(p)>0, r_{-}(p) >0\}\,, \\
	U_{\text{IV}} &= \{ q \in M \;|\; r_{+}(p)<0, r_{-}(p) >0\}\,.
    \end{split}
    \label{eq:regions}
\end{equation}
In the above $p \in \Orthframe(M)$ is any Lorentz orthonormal frame at
$q\in M$.

By hypothesis, our space-time manifold $N$ has a region where it is
asymptotically like Minkowski space.  In such a region a light ray can
go radially inward $(u^{+},u^{-})=(0,1)$ or radially outward
$(u^{+},u^{-})=(1,0)$.  In that asymptotically Minkowski region we can
choose a $p\in \Orthframe(M)$ with the property that $r_{+}(p)>0$ and
$r_{-}(p) <0$ and thus we conclude that $U_{\text{I}} \neq \emptyset$ and
that the familiar asymptotic exterior lies in $U_{\text{I}}$.  According to
\eqref{eq:soln-ODE-r}, an inward future directed radial null geodesic
will have $r(\lambda) = r_{p} + \lambda r_{-}(p)$.  Two important
observations are that for finite positive affine parameter the light
ray will cross $r=\RS$ and in finite affine parameter it will also hit
$r=0$.  This last observation says that our space
may have a singularity because the Cartan structural equations have a
singularity at $r=0$.  We will not address the question of 
whether this is a real or a removable singularity. We will 
concentrate on what happens to null geodesics at $r=\RS$.

\subsection{Schwarzschild Geometry without Coordinates}
\label{sec:Sch-geom}

The key to understanding the geometry of the Schwarzschild solution is
to understand the level sets of the radius function $r: M \to
\bbR_{+}$.  For all practical purposes, both physical and
mathematical, we can take $M$ to be simply connected.  Topology tells
us that $M$ has a universal simply connected cover
$\kappa:\widetilde{M} \to M$.  We have a fiber bundle $\pi: N \to M$.
We can use the covering map $\kappa$ to obtain the pull back bundle
$\tilde{\pi}: \widetilde{N} \to \widetilde{M}$ and we can pull back
all metrics.  The conclusion is that we might as well as well assume
that $M$ is simply connected.  We do this now.

An important ingredient in our discussion is that we can use $r$ as a
Morse function\footnote{There is an application of Morse theory to
black holes by Carter~\cite[p. 187]{Carter:blackholes} but it is different from
ours.} to learn about $M$.  The Einstein equation \eqref{eq:BH-rij}
tells that that the critical points of $r$ are non-degenerate.  Let's
briefly review the argument.  Assume I have a smooth function $f: X
\to \bbR$ where $X$ is a manifold.  The point $p \in X$ is a critical
point if $df\rvert_{p} =0$.  The hessian of $f$ at $x_{0}$ can be
defined intrinsically but it is easier to do it in terms of local
coordinates.  Let $(x^{i})$ and $(y^{i})$ be two local coordinate
systems.  We observe that the matrix of second derivatives has a
non-tensorial transformation law
\begin{equation*}
    \frac{\partial^{2} f}{\partial x^{i}\,\partial x^{j}} =
    \frac{\partial y^{k}}{\partial x^{i}}\, \frac{\partial 
    y^{l}}{\partial x^{j}}\, \frac{\partial^{2}  f}{\partial y^{k} \,
    \partial y^{l}} + \frac{\partial^{2} y^{k}}{\partial x^{i} \, 
    \partial x^{j}}\, \frac{\partial f}{\partial  y^{k}}\,,
\end{equation*}
except at a critical point $p$ where $\partial f/\partial y (p)=0$ and
the above reduces to
\begin{equation*}
    \left.  \frac{\partial^{2} f}{\partial x^{i}\,\partial x^{j}}
    \right\rvert_{p} =  
    \left. \frac{\partial y^{k}}{\partial x^{i}}\right\rvert_{p}\, 
    \left. \frac{\partial y^{l}}{\partial x^{j}}\right\rvert_{p}\,
    \left. \frac{\partial^{2} f}{\partial y^{k} \, \partial y^{l}}
    \right\rvert_{p}.
\end{equation*}
The next thing we observe is that the hessian at a critical point is
given by the second covariant derivative with respect to any
connection $\Gamma$.  The reason is that
\begin{equation*}
    (D_{i}D_{j} f)(p) = \frac{\partial^{2} f}{\partial
    x^{i}\,\partial x^{j}}(p) - \Gamma^{k}{}_{ij}(p) \frac{\partial 
    f}{\partial  x^{k}}(p) = \frac{\partial^{2} f}{\partial
    x^{i}\,\partial x^{j}}(p)\,.
\end{equation*}
If $r:M \to \bbR_{+}$ has critical points then they must be
non-degenerate because of \eqref{eq:BH-rij}.

\begin{figure}[tbp]
    \centering
    \includegraphics[width=0.5\textwidth]{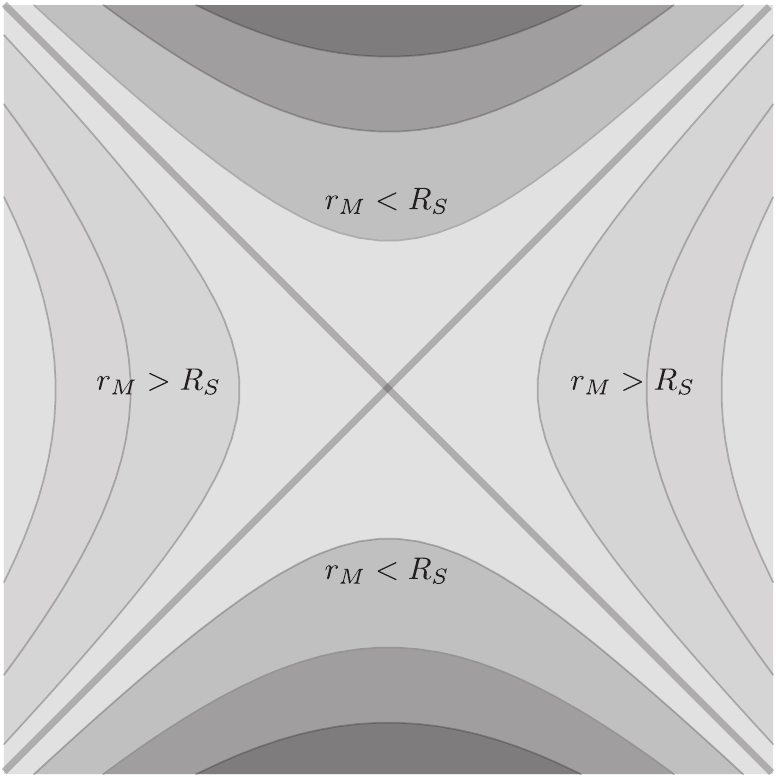}
    \begin{quote}
	    \caption[x]{Behavior of the function $r_{M}$ near the
	    critical point.}
	    \label{fig:critical}
    \end{quote}
\end{figure}

Assume the radius function $r: M \to \bbR_{+}$ has a critical point at
$p \in M$.  We know by \eqref{eq:BH-rij} that this critical point is
non-degenerate.  We also know by Morse's lemma~\cite{Milnor:morse} that
in a neighborhood of $p$ we can find local coordinate $(y^{0},y^{1})$
centered at $p$ that are Minkowski orthonormal at $p$ such that in the
neighborhood we have that
\begin{equation}
    r(y) = \RS + \frac{-(y^{0})^{2} + (y^{1})^{2}}{4\RS}\,.
    \label{eq:r-Morse}
\end{equation}
The neighborhood of any critical point of the function $r_{M}$ looks
like Figure~\ref{fig:critical}.

We begin
analyzing the properties of $r_{M}: M \to \bbR_{+}$.  Let $U_{>} = \{
p \in M \;|\; r_{M}(p)>\RS\}$ and let $U_{<} = \{ p \in M \;|\;
0<r_{M}(p)<\RS\}$.  Note that $U_{>}$ and $U_{<}$ are both open
subsets of $M$.  It is clear from \eqref{eq:BH-dr2} that $dr_{M}(p)
\neq 0$ if $p \in U_{<} \cup U_{>}$.  The implicit function theorem
tells us that the level sets of the function $r_{M}$ give a good
foliation on $U_{<} \cup U_{>}$.  The only question remains what
happens at $r_{M}=\RS$ where we note that $\lVert dr \rVert_{M}^{2} =
0$ by \eqref{eq:BH-dr2}.  Since the metric is Minkowski we
\emph{cannot} conclude that $dr=0$, but we do know that if there is a
critical point then it must be non-degenerate and that it must have 
$r_{M}=R_{S}$.

Next we establish that $r$ must have a critical point.  We choose a
point $p_{1}\in U_{\text{I}} \subset M$ that is in the asymptotic
Minkowski region $r_{M} \gg \RS$ where $r_{+}(q_{1})>0$ and
$r_{-}(q_{1}) <0$ and let $q_{1}\in \Orthframe(M)$ be a Lorentz frame
at $p_{1}$.  We will construct a null broken horizontal curve
beginning at $q_{1}$ that takes us to the critical point.  Begin with
an inward null horizontal curve with initial data $(u^{+},u^{-})=
(0,1)$.  Inserting into \eqref{eq:soln-ODE-r} we see that $r(\lambda)
= r_{q_{1}} + \lambda r_{-}(q_{1})$ and therefore the horizontal curve
arrives to a point $q_{2}\in \Orthframe(M)$ where the sphere radius is
the Schwarzschild radius at
$\lambda_{\RS}=(\RS-r_{q_{1}})/r_{-}(q_{1})>0$.  Note that according
to \eqref{eq:dr-evolution} we have that
\begin{equation*}
    \begin{split}
	\frac{dr_{+}}{d\lambda} &= -\frac{\RS}{4 \left(r_{q_{1}} +
	\lambda r_{-}(q_{1})\right)^{2}}\,, \\
	\frac{dr_{-}}{d\lambda} &=0\,. \\
    \end{split}
\end{equation*}
Thus $r_{-}$ is constant along this horizontal curve and we have 
$r_{-}(q_{2})=r_{-}(q_{1})<0$. We know that at $r=\RS$ we have that 
$\lVert dr \rVert_{M}^{2}=0$ and therefore $r_{+}(q_{2})=0$. We can 
verify this explicitly. Solving the ODE for  $r_{+}$ we see 
that
\begin{equation*}
    r_{+}(\lambda) = r_{+}(q_{1}) -\frac{\RS}{4 r_{-}(q_{1})r_{q_{1}}}
    + \frac{\RS}{4 r_{-}(q_{1})\left(r_{q_{1}} + \lambda
    r_{-}(q_{1})\right)}\,.
\end{equation*}
Inserting $\lambda=\lambda_{\RS}$ and doing some algebra we find the 
desired result.

At the point $q_{2}$ where $r_{-}(q_{2})<0$ and $r_{+}(q_{2})=0$ we
begin a new horizontal curve with initial velocity $(u^{+},u^{-}) =
(-1,0)$. Along this curve we have $r(\lambda) = \RS$ is constant and
\begin{equation*}
    \begin{split}
	\frac{d r_{+}}{d \lambda} &=  
	0\,, \\
	\frac{d r_{-}}{d \lambda} &=
	+\frac{1}{4\RS}\,.
    \end{split}
\end{equation*}
Thus $r_{+}=0$ along this curve and $r_{-}(\lambda) = r_{-}(q_{2}) + 
\lambda/4\RS$. Thus in finite positive $\lambda$ we will get to a 
point  $q_{*}$ where $r_{-}(q_{*})=0$. This is the critical 
point of $r$ that we sought.

We have established that if $M$ has an asymptotic Minkowski region
then there exists a critical point of the radius function $r_{M}$.
Can there be more than one critical point?  The answer is no under our
hypotheses.  The pictorial topological argument is that near each
critical point we have a situation that looks like
Figure~\ref{fig:critical}.  It is very hard to see how two copies of
the figure can be put together consistently.  You can also give a more
analytical argument that has two parts.

The first part is essentially running the proof of the existence
backwards.  Namely we observe that if we start at a critical point
$p_{*}\in M$ of $r_{M}$ then $r_{M}(p_{*})=\RS$ and there are two null
geodesics emanating from $p_{*}$ and along each we have that
$r_{M}=\RS$, see Figure~\ref{fig:critical}.  Lift the geodesics to
horizontal curves.  Along the first horizontal curve we have that
$r_{+}=0$ and $r_{-}$ is a strictly monotonic function of the affine
parameter, and along the other horizontal curve we have the opposite:
$r_{-}=0$ and $r_{+}$ is a strictly monotonic function of the affine
parameter.  This immediately tells us that we cannot have another
critical point along the null geodesics emanating from $p_{*}$.

\begin{figure}[tbp]
    \centering
    \includegraphics[height=0.2\textheight]{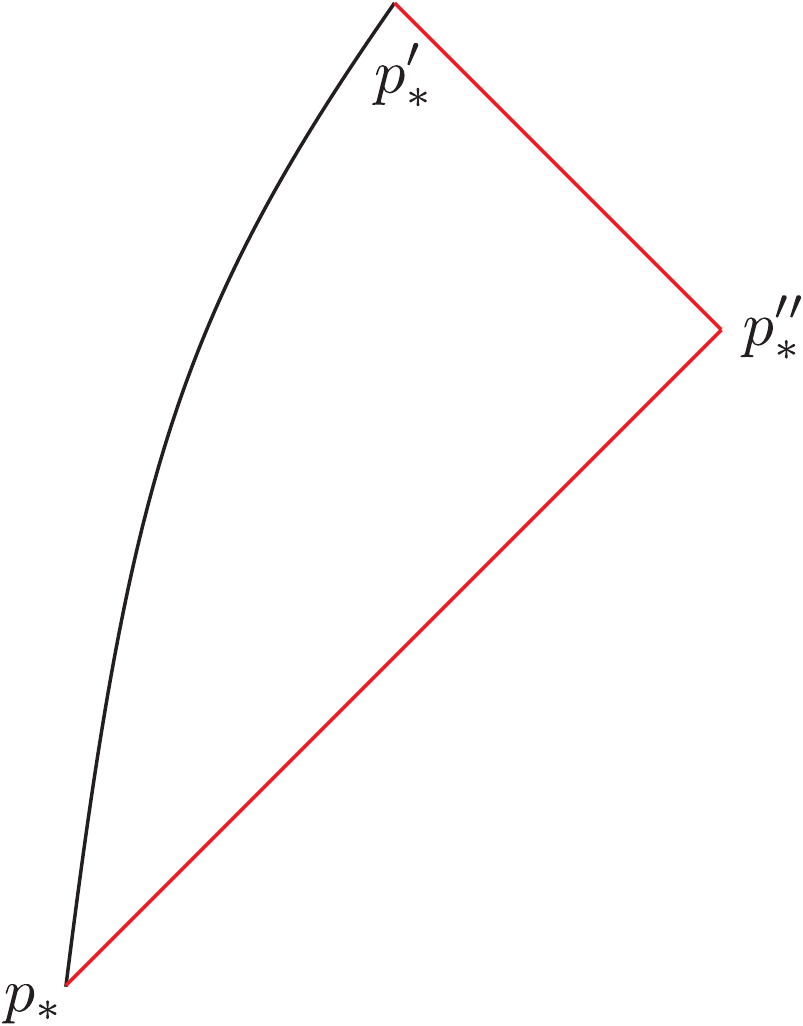}
    \begin{quote}
	    \caption[xx]{Two nearby timelike separated critical points
	    $p_{*}$ and $p'_{*}$ connected by a path and also by a
	    broken null geodesic.}
	    \label{fig:intuition}
    \end{quote}
\end{figure}
The second part of the argument is a bit more involved.  It is proof
by contradiction.  We develop the intuition by studying the case where
we assume that there are nearby timelike separated critical points
$p_{*}$ and $p'_{*}$ as in Figure~\ref{fig:intuition}.  We begin with
a horizontal null curve with initial tangent vector $(u^{+},u^{-}) =
(1,0)$ at $q_{*} \in \Orthframe(M)$ over $p_{*}\in M$.  We evolve the
curve until it reaches a point $q''_{*}$ over $p''_{*}$.  According to
\eqref{eq:dr-evolution} we have that $r_{+}(q''_{*})=0$ and
$r_{-}(q''_{*}) <0$.  Next we begin a null curve with initial tangent
vector $(u^{+},u^{-}) = (0,1)$ that will take us to a point $q'_{*}$
over the critical point $p'_{*}$.  But according to
\eqref{eq:dr-evolution} we have that $r_{-}(q'_{*}) = r_{-}(q''_{*})
<0$.  This contradicts $dr(q')=0$.  More generally, assume that
you have two critical points on $M$ with $p'_{*}$ in the future of
$p_{*}$.
\begin{figure}[tbp]
    \centering
    \includegraphics[height=0.2\textheight]{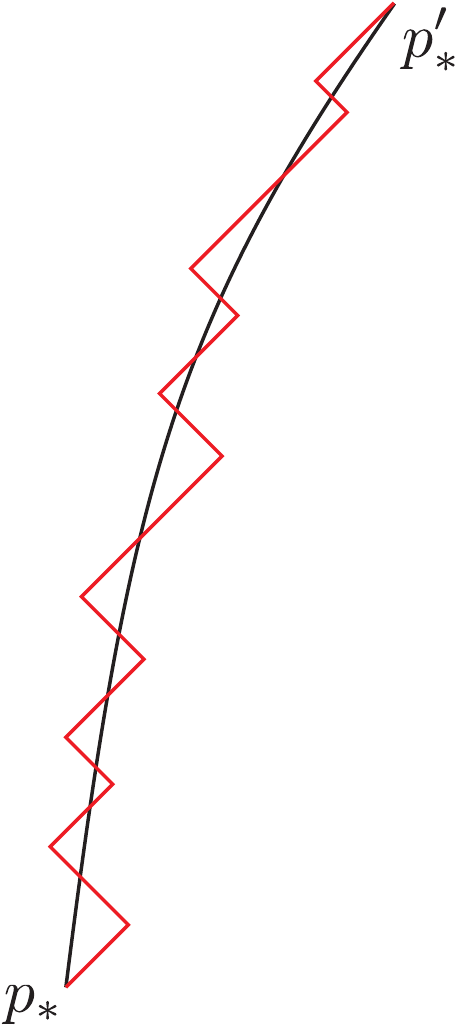}
    \begin{quote}
	    \caption[xxx]{Two causally separated critical points $p_{*}$
	    and $p'_{*}$ connected by a causal path and also by a
	    broken null geodesic.}
	    \label{fig:timelike}
    \end{quote}
\end{figure}
Since $M$ is connected there is a causal curve between them.  We
approximate the causal curve by the zig-zag path of null geodesics as
in Figure~\ref{fig:timelike}.  If we apply the previous argument piece
by piece to the zig-zag we conclude that $dr(q'_{*}) \neq 0$. The 
argument can be extended to the case where the conjectured critical 
points are not causally connected by using broken null geodesics that 
are future and past directed.

\subsection{The Kruskal Spacetime}
\label{sec:kruskal}

\begin{figure}[tbp]
    \centering
    \includegraphics[width=0.5\textwidth]{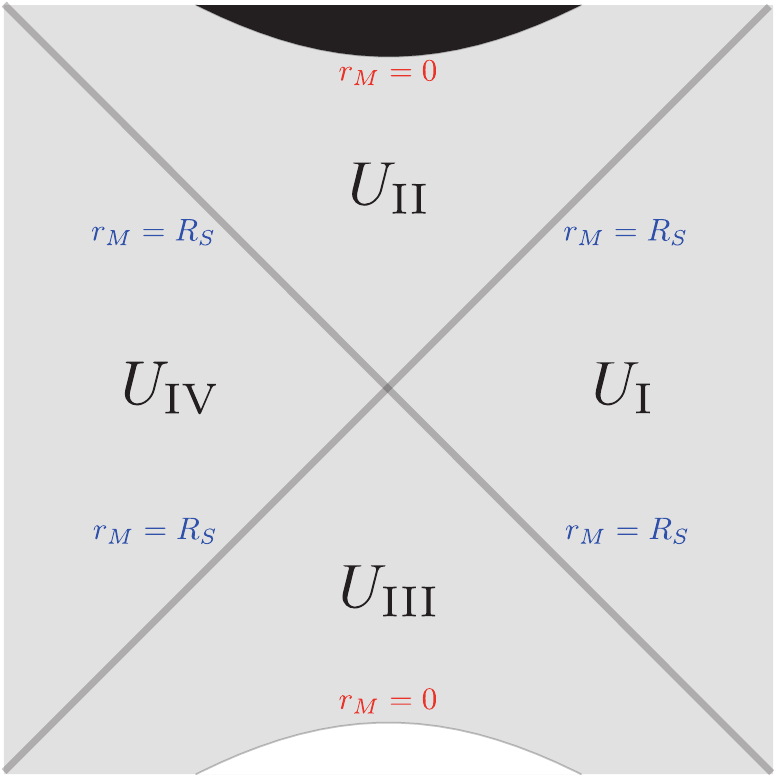}
    \begin{quote}
	    \caption[xxxs]{The four regions of the Kruskal spacetime
	    with the black hole and the white hole singularities as
	    indicated.}
	    \label{fig:kruskal}
    \end{quote}
\end{figure}

Since there is only one critical point we have the standard Kruskal 
diagram, see Figure~\ref{fig:kruskal}, of the Schwarzschild geometry  with 
the four regions associated with \eqref{eq:regions}.
The asymptotic Minkowski region is in Region $U_{\text{I}}$.  In
Region $U_{\text{II}}$ we have that $r_{\pm}<0$ and
\eqref{eq:soln-ODE-r} tells us that all future directed null geodesics
end up at $r_{M}=0$ in finite affine parameter.  Therefore light rays
cannot escape Region $U_{\text{II}}$.  This is the black hole region.
Region $U_{\text{III}}$ is the white hole region which is the time
reversal image of the black hole region.  Region $U_{\text{IV}}$ is
the parity image of $U_{\text{I}}$.  The two Minkowski like regions
are causally disconnected.  The properties of the radial null
geodesics in the various regions are easily determined using
\eqref{eq:soln-ODE-r}.

Note that the Killing vector field $X$ is a null vector field on the
null lines defined by $r_{M}=\RS$ in the Kruskal spacetime.  If we
think in terms of the fibration $\pi: N \to M$.  The fiber over where
the two line intersects is called the the bifurcation $2$-sphere.  The
fibers over the two null geodesics at $r_{M}=\RS$ define the bifurcate
Killing horizon.

There are real singularities at $r_{M}=0$. We do not get any new 
insight into the nature of the singularities using these methods. For 
this reason we will not say anything about it.

\subsection{Schwarzschild Coordinates}
\label{sec:sch-coord}

We conclude by using our geometrical data to write down the metric in
standard Schwarzschild coordinates.  On the open set $V \subset
\Orthframe(M)$ that is the complement to the closed set $r^{-1}(\RS)
\subset \Orthframe(M)$ we can write
\begin{equation}
    \begin{split}
	\tau &= \frac{r_{1}}{1-\RS/r}\; \theta^{0} + \frac{r_{0}}{1-\RS/r}\; 
	\theta^{1}\,,\\
	dr & = r_{0}\theta^{0} + r_{1} \theta^{1}\,.
    \end{split}
    \label{eq:tau-dr}
\end{equation}
The inverse relationship is 
\begin{equation}
    \begin{split}
	\theta^{0} &= -r_{1}\,\tau - \frac{r_{0}}{1-\RS/r}\;dr\,,\\
	\theta^{1} & = -r_{0}\tau + \frac{r_{1}}{1-\RS/r} \;dr\,.
    \end{split}
    \label{eq:tau-dr-theta}
\end{equation}
From this we learn that on $V$ we have
\begin{equation}
    -\left(\theta^{0} \right)^{2} + \left(\theta^{1} \right)^{2} =
    -\left(1-\RS/r\right) \tau^{2} + \left(1-\RS/r\right)^{-1} 
    (dr)^{2}\,.
    \label{eq:Schw-metric}
\end{equation}
which is the metric in standard Schwarzschild coordinates because
$\tau$ is a closed $1$-form and therefore locally exact\footnote{There
are really four different functions
$t_{\text{I}},t_{\text{II}},t_{\text{III}},t_{\text{IV}}$
corresponding to the four regions
$U_{\text{I}},U_{\text{II}},U_{\text{III}},U_{\text{IV}}$ that make up
$V$.}, $\tau =dt$.  By taking a section you can pull these structures
back to the base $M$.

\subsection{Redshift without Coordinates}
\label{sec:redshift}

This discussion is treated in standard texts.
We have a timelike Killing vector field $X$ in regions I and IV and we
restrict to observers in either of these regions.  Let $k$ be a
tangent vector to a null geodesic, \emph{i.e.}, $D_{k}k=0$ and $\lVert
k \rVert^{2}=0$.  It is elementary to show that $D_{k}(X \cdot k)=0$.
In other words $X\cdot k$ is constant along the geodesic.  We know
that if $k$ is the wave vector of a beam of light then an observer at $q
\in N$ with (timelike) $4$-velocity $u$ will measure the frequency to
be $\omega(q) = - u \cdot k$.  Consider two observers $E$ and $O$ at
fixed radii $r_{E}$ and $r_{O}$.  A photon is emitted by $E$ and
observed by $O$.  We note that $u_{E} = (1-\RS/r_{E})^{-1/2} X_{E}$
and that $u_{O} = (1-\RS/r_{O})^{-1/2} X_{O}$ because the observers
are at fixed radii.  Using the constancy of $X \cdot k$ along the null
geodesic we conclude that
\begin{equation}
    \frac{\omega_{O}}{\omega_{E}} = 
    \sqrt{\frac{1-\RS/r_{E}}{1-\RS/r_{O}}}\,.
    \label{eq:redshift}
\end{equation}

\section*{Acknowledgments}%
\label{sec:acknowledgments}%
\addcontentsline{toc}{section}{Acknowledgments}%

I would like to thank G.~Galloway for reading an earlier draft of the
manuscript.  I would also like to thank R.~Bryant who explained to me
some of the intricacies of the method of moving frames about 15 years
ago.  Over the years, I.M.~Singer has generously explained and shared
his geometrical insights about the theory and the importance of
principal bundles.  I would like to also thank him for reading an
earlier draft.  This work was supported in part by National Science
Foundation grants PHY--0244261 and PHY--0554821.

%
\appendix

\section{Cartan's Lemma}
\label{sec:cartan-lemma}

Cartan's lemma is the observation that if $\{\varphi^{i}\}$ is a linearly 
independent collection of $1$-forms and if \{$\alpha_{i}\}$ are $1$-forms 
such that $\alpha_{i}\wedge \varphi^{i}=0$ then there exists 
coefficients $\{a_{ij}\}$ with $a_{ij}=a_{ji}$ such that $\alpha_{i} = 
a_{ij}\varphi^{j}$.

A corollary to Cartan's lemma is the statement that if you have a
collection of $1$-forms $\{\beta_{ij}\}$ with $\beta_{ij}=-\beta_{ji}$ and
if $\beta_{ij}\wedge\varphi^{j} =0$ then $\beta_{ij}=0$.  To prove
this we note that Cartan's lemma implies that there exists
coefficients $b_{ijk} =b_{ikj}$ such that $\beta_{ij} =
b_{ijk}\varphi^{k}$.  But $b_{ijk}$ is skew symmetric under $i
\leftrightarrow j$ but symmetric under $j \leftrightarrow k$ and
therefore $b_{ijk}=0$.  This corollary is responsible for the
uniqueness of the Levi-Civita connection, \emph{i.e.}, the fundamental
lemma of riemannian geometry.

\section{Lightcone Conventions}
\label{sec:conventions}
\begin{align}
   \theta^{\pm} & = \theta^{0} \pm \theta^{1} &
   \frac{\partial}{\partial \theta^{\pm}} & = \half \left(
   \frac{\partial}{\partial \theta^{0}} \pm
   \frac{\partial}{\partial \theta^{1}} \right)
   \label{eq:conv-1} \\
    ds^{2} &= -(\theta^{0})^{2} + (\theta^{1})^{2}\,, & ds^{2}   & = 
    -\half \left( \theta^{+} \otimes \theta^{-} + 
    \theta^{-}\otimes \theta^{+} \right)
    \label{eq:conv-2}  \\
    \eta_{+-} & = -\half & \eta^{+-} & = -2\,,
    \label{eq:conv-3}  \\
    \theta^{0} \wedge \theta^{1} & = -\half \theta^{+} \wedge
    \theta^{-}\,, & \epsilon_{01}=+1, &\quad \epsilon^{01} = -1\,,
    \label{eq:conv-4} \\
    \epsilon_{+-}=-\half, &\quad \epsilon^{+-} = +2\,,
    & \epsilon^{-}{}_{-} = +1, &\quad \epsilon^{+}{}_{+} = -1\,,
    \label{eq:conv-5}\\
    v^{+} = -2v_{-}\,, &\quad  v^{-} = -2 v_{+}\,, & v_{-} = -\half 
    v^{+}\,, &\quad v_{+} = -\half v^{-}\,,
    \label{eq:conv-6} \\
    \lVert v \rVert^{2}  = -v^{+}v^{-} &= -4 v_{+}v_{-}\,,
     & v^{+}v_{+} = v^{-}v_{-} &= \half \lVert v \rVert^{2}\,,
    \label{eq:conv-7} \\
    \Box f  = \eta^{ij}f_{;ij} &= -4 f_{;+-}\,,
\end{align}

\section{Cartan's Approach to Geodesics}
\label{sec:geodesics-cartan}

Cartan studies geodesics on a manifold $N$ by using the structural
equations to study horizontal curves in the bundle of
frames~\cite{Cartan:riemann}.  In fact, Cartan often studies families
of geodesics via the exponential map generalized to the bundle of
frames.

Let $\pi:\Orthframe(N) \to N$ be the orthonormal frame bundle with
canonical coframing $(\theta^{\mu}, \omega_{\mu\nu})$.  It is well
known that if $q \in \Orthframe(N)$ with $p=\pi(q) \in N$ then a curve
in $N$ based at $p$ uniquely lifts to a horizontal curve in
$\Orthframe(N)$ beginning at $q$.  Using his structural equations,
Cartan sets up a system of ordinary differential equations satisfied
by the horizontal lift of the geodesic.  Cartan considers a map $E:
\bbR \times \bbE^{n} \to \Orthframe(N)$.  Fix $q\in N$ and $u \in
\bbE^{n}$ then we have that $E(0,u)=q$ and as $\lambda$ varies we have
that $E(\lambda,u)$ will be the horizontal curve with ``constant
velocity'' $u$:
\begin{equation}
    \begin{split}
	\theta^{\mu}\left( E_{*}\left(\frac{\partial}{\partial 
	\lambda} \right) \right) & = u^{\mu}\;, \quad(\text{constant 
	velocity})\\
	\omega_{\mu\nu}\left( E_{*}\left(\frac{\partial}{\partial 
	\lambda} \right) \right) & =0\;. \quad(\text{horizontal})	
    \end{split}
    \label{eq:C-geodesic-0}
\end{equation}
The dual versions of these statements are
\begin{equation}
    \begin{split}
	E^{*} \theta^{\mu} &= u^{\mu}\,d\lambda + \vartheta^{\mu}\,, \\
	E^{*}\omega_{\mu\nu} &= \varpi_{\mu\nu}\,,
    \end{split}
    \label{eq:C-geodesic}
\end{equation}
where $\vartheta^{\mu}$ and $\varpi_{\mu\nu}$ are unknown $1$-forms on
$\bbR\times \bbE^{n}$ that are independent of $d\lambda$.  Note that on
$\bbR \times \bbE^{n}$ there are natural global cartesian coordinates
$(\lambda,u)$ and differential forms can be assigned a bi-degree $(k,l)$
where $k=0,1$ and $l=0,1,\ldots,n$.  For example, $d\lambda$ has bi-degree
$(1,0)$ and $\vartheta^{\mu}$ has bi-degree $(0,1)$.
\begin{quotation}
    \small
\begin{comment}
    \label{comment:cart}
    If $N=\bbE^{n}$ with cartesian coordinates $x$ then the map $\pi
    \circ E$ is given by $(\lambda,u) \mapsto x^{\mu}=\lambda
    u^{\mu}$.  Note that $dx^{\mu} = u^{\mu}\;d\lambda + \lambda\;du^{\nu}$
    and in comparing with \eqref{eq:C-geodesic} we see that
    $\vartheta^{\mu} = \lambda\; du^{\mu}$.
\end{comment}
\begin{comment}
    On the vector space $\bbR^{k}$ with standard coordinates
    $(x^{1},\ldots,x^{k})$, the exterior derivative $d$ acting on a
    $p$-form $\alpha = \alpha_{J} dx^{J}$, $\lvert J \rvert =p$
    (multi-index notation) is simply given by
    \begin{equation*}
        d\alpha = dx^{i} \wedge \frac{\partial \alpha}{\partial x^{i}}\,,
    \end{equation*}
    where
    \begin{equation*}
	\frac{\partial \alpha}{\partial x^{i}} = \frac{\partial 
	\alpha_{J}}{\partial x^{i}}\; dx^{J}\,.
    \end{equation*}
\end{comment}
\end{quotation}

Taking the exterior derivatives of \eqref{eq:C-geodesic} and using 
the comments we have
\begin{equation}
    \begin{split}
	-\varpi_{\mu\nu} \wedge (u^{\nu}\,d\lambda + \vartheta^{\nu}) &=
	du^{\mu} \wedge d\lambda + d\lambda \wedge
	\frac{\partial\vartheta^{\mu}}{\partial \lambda} + du^{\nu} \wedge
	\frac{\partial\vartheta^{\mu}}{\partial u^{\nu}}\,, \\
	E^{*}\left(-\omega_{\mu\kappa} \wedge
	\omega_{\kappa\nu} +\half R_{\mu\nu\rho\sigma} \theta^{\rho} 
	\wedge \theta^{\sigma} \right) &= d\lambda \wedge
	\frac{\partial\varpi_{\mu\nu}}{\partial \lambda} + du^{\lambda}
	\wedge \frac{\partial \varpi_{\mu\nu}}{\partial
	u^{\lambda}}\,,
    \end{split}
    \label{eq:C-geodesic-1}
\end{equation}
Identifying the terms that have bi-degree $(1,1)$ we find
\begin{equation}
    \begin{split}
        \frac{\partial\vartheta^{\mu}}{\partial \lambda} &=
	du^{\mu} + \varpi_{\mu\nu} u^{\nu}\,,\\
        \frac{\partial\varpi_{\mu\nu}}{\partial \lambda} &=
	r_{\mu\nu\rho\sigma}\, u^{\rho}\,\vartheta^{\sigma}\,,\\
    \end{split}
    \label{eq:C-geodesic-2}
\end{equation}
where $r_{\mu\nu\rho\sigma} = E^{*}R_{\mu\nu\rho\sigma}$.  The initial
conditions on these differential equations are $\vartheta^{\mu}
\rvert_{\lambda=0}=0$ and $\varpi_{\mu\nu} \rvert_{\lambda=0}=0$.
This follows from the condition that $E(0,u)=q$ for all $u$, see for 
example Comment~\ref{comment:cart}.

Equations~\eqref{eq:C-geodesic-2} can be combined into a second order 
differential equation
\begin{equation}
    \frac{\partial^{2}\vartheta^{\mu}}{\partial \lambda^{2}} = 
    r_{\mu\nu\rho\sigma}\, u^{\nu} u^{\rho}\, \vartheta^{\sigma}\,,
    \label{eq:C-geodesic-3}
\end{equation}
with initial conditions $\vartheta^{\mu}\rvert_{\lambda=0}=0$ and
$\left.  \left(\partial\theta^{\mu}/\partial\lambda\right)
\right\rvert_{\lambda=0} =du^{\mu}$.  This  equation is Cartan's
equation for a Jacobi vector field.  It tells you how the coframe
changes along the horizontal lift of a geodesic.

\subsection{Holonomy and Symmetric Spaces}
\label{sec:holonomy}

Using the methods of the previous section it is easy to understand the
basic properties of symmetric spaces.  Choose a point $q \in
\Orthframe(M)$, let $\Gamma_{q}$ be the holonomy group at $q$ and let
$\Phi_{q} \subset \Orthframe(M)$ be the set of all points in the frame
bundle that are connected to $q$ by a piecewise differentiable
horizontal curve.  The basic theorem is that the holonomy bundle
$\Phi_{q}$ is a sub-bundle of $\Orthframe(M)$ with structure group
$\Gamma_{q} \subset \SOrth(n)$ and that $M=\Phi_{q}/\Gamma_{q}$,
see~\cite{KN:I}.

Next we show that the riemannian connection restricted to the 
holonomy bundle is a $\Gamma_{q}$-connection. To do this we write the 
Lie algebra 
\begin{equation}
    \so(n) = \lieg \oplus \lieh 
    \label{eq:Lie-decomp}
\end{equation}
where $\lieg$ is the Lie 
algebra of $\Gamma_{q}$ and $\lieh$ is a complementary subspace. Under 
this decomposition the Cartan structural equations become
\begin{equation}
    \begin{split}
	d\omega^{\alpha} &= 
	-\half f^{\alpha}{}_{\beta\gamma} \omega^{\beta}\wedge \omega^{\gamma}
	- f^{\alpha}{}_{\beta c} \omega^{\beta}\wedge \omega^{c}
	-\half f^{\alpha}{}_{bc} \omega^{b}\wedge \omega^{c}
	+ \half R^{\alpha}{}_{\mu\nu} \theta^{\mu} \wedge \theta^{\nu}\,,\\
	d\omega^{a} &= 
	-\half f^{a}{}_{\beta\gamma} \omega^{\beta}\wedge \omega^{\gamma}
	- f^{a}{}_{\beta c} \omega^{\beta}\wedge \omega^{c}
	-\half f^{a}{}_{bc} \omega^{b}\wedge \omega^{c}
	+ \half R^{a}{}_{\mu\nu} \theta^{\mu} \wedge \theta^{\nu}\,,\\
	d\theta^{\mu} &= -A^{\mu}{}_{\nu\alpha} \omega^{\alpha}\wedge \theta^{\nu}
	-A^{\mu}{}_{\nu a} \omega^{a}\wedge \theta^{\nu}.
    \end{split}
    \label{eq:c-split}
\end{equation}
In the above the indices $a,b,c$ refer to $\lieg$ and
$\alpha,\beta,\gamma$ refer to $\lieh$.  The $f$ are the structure
constants for $\so(n)$ adapted to the decomposition
\eqref{eq:Lie-decomp} and the $A$ are other constants associated to
the same decomposition\footnote{These are really associated with the
decomposition of the basic representation of $\so(n)$ in terms of the
decomposition \eqref{eq:Lie-decomp}.}.  The holonomy bundle is a
sub-bundle that solves the equation $\omega^{\alpha}=0$.  The reason
is that a basis for $T_{q}\Phi_{q}$ is $\{e_{\mu}\} \cup \{e_{a}\}$,
\emph{i.e.}, need the horizontal curves that are used to construct the
holonomy bundle and also need the holonomy Lie algebra.  The Frobenius
theorem requires $f^{\alpha}{}_{bc}=0$ and $R^{\alpha}{}_{\mu\nu}=0$
for an integrable distribution.  This means that $\lieg$ is a
subalgebra of $\so(n)$ as required and there is no curvature in the
``$\lieh$-direction''.  Restricting to the holonomy sub-bundle
$\Phi_{q}$ we have structural equations
\begin{equation}
    \begin{split}
	d\omega^{a} &= 
	-\half f^{a}{}_{bc} \omega^{b}\wedge \omega^{c}
	+ \half R^{a}{}_{\mu\nu} \theta^{\mu} \wedge \theta^{\nu}\,,\\
	d\theta^{\mu} &= 
	-A^{\mu}{}_{\nu a} \omega^{a}\wedge \theta^{\nu}.
    \end{split}
    \label{eq:c-split-1}
\end{equation}
These equation tell us that the restriction of the connection to the
holonomy bundle is a $\Gamma_{q}$-connection\footnote{These arguments
generalize to the holonomy bundle of a generic bundle not just the
frame bundle.}.

How do symmetric spaces arise from this viewpoint\footnote{I ignore
some issues of connected component, etc.}?  If the curvature is
covariantly constant then it is a constant function on the holonomy
sub-bundle.  The reason is that $dR= -\omega\cdot R +
(\nabla_{\lambda}R)\theta^{\lambda}=-\omega\cdot R$ which vanishes
along a horizontal curve.  Consequently $R_{\mu\nu\rho\sigma}$ must be
a constant function on the holonomy sub-bundle $\Phi_{q}$.  This means
that $R^{a}{}_{\mu\nu}$ are constant and therefore
equations~\eqref{eq:c-split-1} are the Maurer-Cartan equations for a
Lie group $G$.  Therefore $\Phi_{q} \approx G$ and $M=G/\Gamma_{q}$.
There is a stronger statement we can make.  The Maurer-Cartan
equations \eqref{eq:c-split-1} admit a symmetry $\omega \to \omega$
and $\theta \to -\theta$.  This is the famous Cartan involution that
leads to symmetric Lie algebras and associated symmetric spaces.  The
reason for ``symmetric'' may be see in \eqref{eq:C-geodesic-3}.  Note
that $r_{\mu\nu\rho\sigma}u^{\nu}u^{\sigma}$ are constant and
therefore $\lambda\to -\lambda$ is a symmetry of the differential
equation.  This means that by integrating \eqref{eq:C-geodesic-3} to
construct the metric we have an isometry between the point at time
$\lambda$ and the one at $-\lambda$.  This is the Cartan local
isometry in a symmetric space\footnote{Note that we can use
\eqref{eq:C-geodesic-3} to conclude the converse.  If we have the
local isometry about any point then \eqref{eq:C-geodesic-3} must be
even under $\lambda\to -\lambda$ and therefore $r_{\mu\nu\rho\sigma}$
must be an even function and therefore the derivative vanishes at
$\lambda=0$.  This being true at all points implies that $\nabla
R=0$.}.

\relax 

%

\providecommand{\href}[2]{#2}\begingroup\raggedright\endgroup

\end{document}

%% file: Ricci.tex
\def\mysavedown#1{\edef\mysubs{\mysubs#1}}
\def\mysaveup#1{\edef\mysups{\mysups#1}}
\def\mydown#1{{\mytensor}_{\vphantom{\mysubs}#1}}
\def\myup#1{{\mytensor}^{\vphantom{\mysups}#1}}
\def\tensor#1#2{
  #1
  \def\mytensor{\vphantom{#1}}
  \def\mysubs{\relax}
  \def\mysups{\relax}
  \let\down=\mysavedown
  \let\up=\mysaveup
  #2
  \let\down=\mydown
  \let\up=\myup
  #2
  }


%% file: blackhole-paper.bbl
\begin{thebibliography}{10}

\bibitem{einstein-equations}
H.~Stephani, D.~Kramer, M.~MacCallum, C.~Hoenselaers, and E.~Herlt, {\em Exact
  solutions of {E}instein's field equations}.
\newblock Cambridge Monographs on Mathematical Physics. Cambridge University
  Press, Cambridge, second~ed., 2003.

\bibitem{carroll:book}
S.~Carroll, {\em Spacetime and geometry: an introduction to general
  relativity}.
\newblock Addison Wesley, 2004.

\bibitem{dinverno:book}
R.~d'Inverno, {\em Introducing {E}instein's relativity}.
\newblock The Clarendon Press, Oxford University Press, New York, 1992.

\bibitem{MTW}
C.~W. Misner, K.~S. Thorne, and J.~A. Wheeler, {\em Gravitation}.
\newblock W. H. Freeman and Co., San Francisco, Calif., 1973.

\bibitem{Hawking-Ellis:book}
S.~W. Hawking and G.~F.~R. Ellis, {\em The large scale structure of
  space-time}.
\newblock Cambridge University Press, London, 1973.
\newblock Cambridge Monographs on Mathematical Physics, No. 1.

\bibitem{Wald:book}
R.~M. Wald, {\em General relativity}.
\newblock University of Chicago Press, Chicago, IL, 1984.

\bibitem{KN:I}
S.~Kobayashi and K.~Nomizu, {\em Foundations of differential geometry. {V}ol
  {I}}.
\newblock Interscience Publishers, a division of John Wiley \& Sons, New
  York-London, 1963.

\bibitem{ONeill:submersion}
B.~O'Neill, ``The fundamental equations of a submersion,'' {\em Michigan Math.
  J.} {\bf 13} (1966) 459--469.

\bibitem{ONeill:SRG}
B.~O'Neill, {\em Semi-{R}iemannian Geometry}.
\newblock Academic Press, 1983.

\bibitem{Warner}
F.~W. Warner, {\em Foundations of Differentiable Manifolds and Lie Groups}.
\newblock Scott, Foresman and Co., 1971.

\bibitem{Cartan:riemann}
{\'E}.~Cartan, {\em Le\c cons sur la {G}\'eom\'etrie des {E}spaces de
  {R}iemann}.
\newblock Gauthier-Villars, Paris, 1946.
\newblock 2d ed.

\bibitem{Carter:blackholes}
B.~Carter, ``Black hole equilibrium states,'' in {\em Black Holes}, C.~DeWitt
  and B.~DeWitt, eds., pp.~57--214.
\newblock Gordon and Breach Science Publishers, 1973.
\newblock Lectures presented at the {Les Houches Summer School} in August~1972.

\bibitem{Milnor:morse}
J.~Milnor, {\em Morse theory}.
\newblock Based on lecture notes by M. Spivak and R. Wells. Annals of
  Mathematics Studies, No. 51. Princeton University Press, Princeton, N.J.,
  1963.

\end{thebibliography}
